\documentclass[conference,compsoc]{IEEEtran}
\ifCLASSOPTIONcompsoc
  \usepackage[nocompress]{cite}
\else
  \usepackage{cite}
\fi
\ifCLASSINFOpdf
 
\else
 
\fi

\usepackage{booktabs} %
\usepackage{graphicx}
\usepackage[hyphens]{url}
\usepackage{hyperref}
\hypersetup{breaklinks=true}
\PassOptionsToPackage{hyphens}{url}
\usepackage{amsmath}
\usepackage{multirow}
\usepackage{mathtools}  %
\usepackage{tikz}
\usepackage[dvipsnames]{xcolor}
\usepackage{framed}   %
\usepackage[most]{tcolorbox}
\usepackage{longtable}
\usepackage{array}
\usepackage{tabularx}
\usepackage{enumitem}
\usepackage{scalerel}

\tcbset{
  insightbox/.style={
    colback=gray!10, %
    colframe=black, %
    boxrule=0.2mm, %
    arc=1mm, %
    left=1mm, %
    right=1mm, %
    top=0mm, %
    bottom=0mm, %
    width=\linewidth, 
  }
}

\newcommand{\blackcircled}[1]{%
    \tikz[baseline=(char.base)]{
        \node[shape=circle, fill=black, text=white, inner sep=0.5 pt] (char) {#1};
    }%
}

\newcommand{\bluecircled}[1]{%
    \tikz[baseline=(char.base)]{
        \node[shape=circle, fill=NavyBlue, text=white, inner sep=0.5 pt] (char) {#1};
    }%
}

\newcommand{\new}[1]{{\color{black}#1}}

\newcommand{\cam}[1]{{\color{black}#1}}

\begin{document}

\title{She Spoofed Sea Ships by the Sea Shore: \new{Measuring Large-Scale GPS Spoofing in Global Maritime Traffic}}

\author{\IEEEauthorblockN{Anna Raymaker, Ryan Von Brock, Ryan Pickren, Animesh Chhotaray, Frank Li, Saman Zonouz, Raheem Beyah}
\IEEEauthorblockA{Georgia Institute of Technology}}

\maketitle

\begin{abstract}
GPS spoofing has emerged as a serious threat to maritime security, yet its global prevalence, persistence, and structure remain largely unmeasured. In this paper, we present the first large-scale measurement study of maritime GPS spoofing, using global Automatic Identification System (AIS) data, which contain the GPS coordinates broadcasted over time by ships across the world. \new{We focus on large-scale regional spoofing, where external interference displaces many vessels across an area at once, leaving a recognizable signature of physically implausible motion correlated across ships; our motion-aware, marine-specific framework identifies this signature and grades the evidence for GPS spoofing in each region it finds.} Applying our approach to AIS data from over 367,000 vessels collected between late November 2024 and early February 2025, we identify 31 persistent \new{anomalous} hotspots across high-traffic maritime regions, \new{at least 22 of which show strong evidence of GPS spoofing,} with spatial and temporal structure aligning with regional conflict and economic sanctions. Notably, our method found that the spoofing activity in the Red Sea responsible for the highly-publicized  grounding of the 75,000-ton container ship, \emph{MSC Antonia}, was ongoing months before the incident, which has not been previously documented. \new{Similarly, we detected persistent spoofing in the Strait of Hormuz over a year before the 2026 Iran war brought commercial shipping through the Strait to near-standstill.} Together, this work establishes GPS spoofing as a widespread, recurring, and measurable threat to global maritime navigation.
\end{abstract}

\IEEEpeerreviewmaketitle

\section{Introduction}
\label{sec:intro}
\begin{figure}[]
  \centering
  \includegraphics[width=0.99\linewidth]{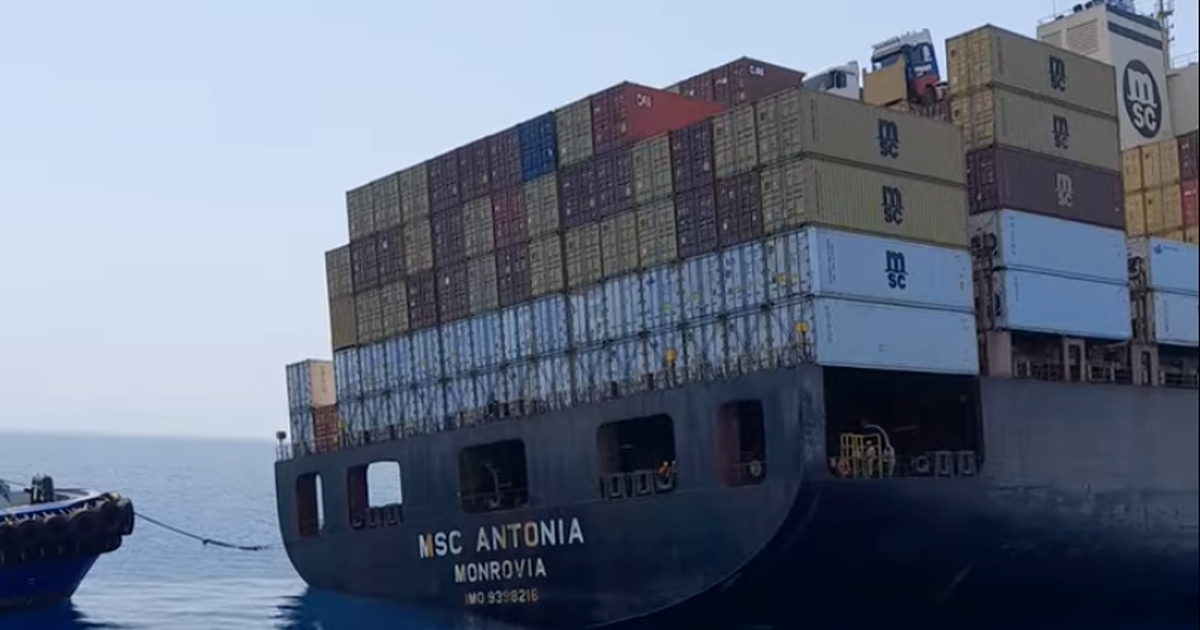}
  \vspace{4pt}
  \includegraphics[width=\linewidth]{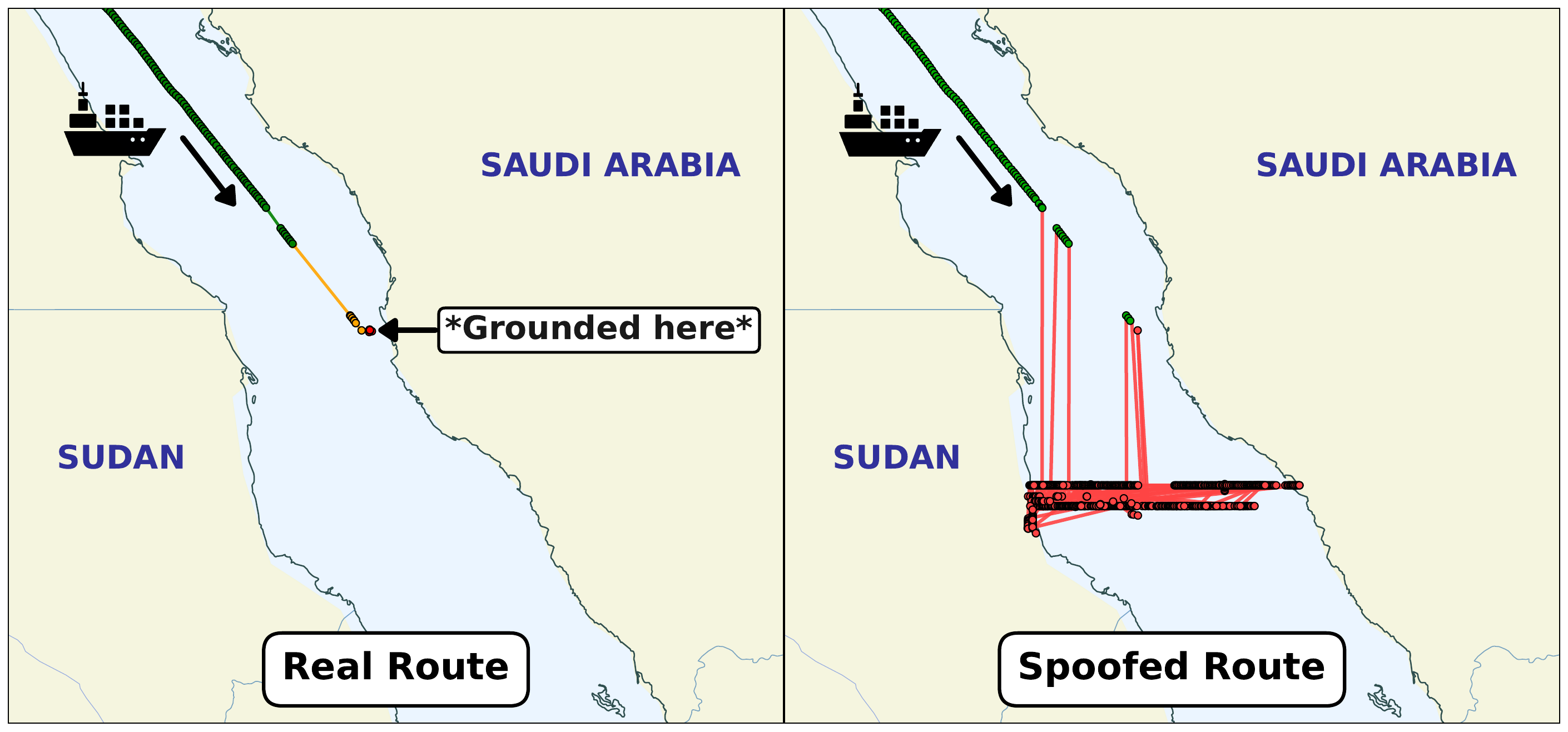}
  \caption{(\textit{Top}) The \emph{MSC Antonia} grounded in the Red Sea in May 2025 after a spoofing incident~\cite{hancock_msc_antonia}.  
  (\textit{Bottom}) AIS comparison showing the vessel’s true route and grounding point (left) versus the spoofed route (right).}  
  \label{fig:antonia_case}
\end{figure}

Maritime systems form an essential part of global infrastructure and play a central role in sustaining the world economy. Around 90\% of global trade is transported by sea~\cite{ics_shipping_data}, and maritime operations also support critical services such as energy transport, logistics, and undersea communication infrastructure. The safety and efficiency of these systems depend on reliable satellite-based positioning and navigation, particularly the Global Positioning System (GPS). However, recent incidents demonstrate that interference with GPS signals can lead to navigational errors, operational disruptions, and physical damage to vessels~\cite{macintyre_gps_spoofing_shipping,editorial_china_gps_spoofing}. In fact, reports found that over 80\% of marine casualties involved human or navigational factors in 2024, underscoring the stakes of degraded positioning~\cite{emsa_2024}. A recent event, shown in Figure~\ref{fig:antonia_case}, is the grounding of the 75,000-ton container ship \emph{MSC Antonia} in the Red Sea, a major incident attributed to GPS spoofing~\cite{hancock_msc_antonia}. The grounding immobilized the vessel for more than five weeks, incurring multi-million-dollar costs in salvage, recovery, and operational disruption~\cite{lloyds_timeline,wecox_antonia,lockton_antonia}.

Such incidents are becoming more common, with reports linking GPS interference to vessel groundings, collisions, and geopolitical tension~\cite{windward_front_eagle,bockmann_uk_tanker,chambers_msc_antonia,woody_navy_accidents}. Interviews with professional mariners further indicate that GPS spoofing is a recurrent operational concern~\cite{raymaker2025sea}. Although these findings confirm a growing threat, the overall scale and distribution of maritime GPS spoofing remain unclear.

Prior research has extensively demonstrated GPS spoofing across automotive and aerial domains~\cite{tippenhauer2011requirements,zeng2018all,sathaye2022experimental,balduzzi2014security,pavur2020tale}, and proposed detection and mitigation techniques in controlled  settings~\cite{liu2021stars,davidovich2022visas,amro2022navigation}. However, these efforts do not measure spoofing prevalence or persistence at global scale, nor do they characterize spoofing using real-world maritime traffic.

To address this gap, we present the first global measurement of \new{large-scale} maritime GPS spoofing using real-world vessel movement data. \new{Our approach starts from a simple question: what would large-scale external GPS spoofing of a region look like in vessel movement data? Because such interference overpowers GPS across an area, it displaces the reported positions of many vessels at once rather than any single ship. This produces a recognizable signature: affected vessels report physically implausible motion, their displacements coincide in space and time, and they concentrate within the bounded region the source covers. These properties are invariants of regional external spoofing, holding regardless of the spoofer's equipment or intent. Our framework is built to identify them: it flags physically implausible motion on each vessel using a Kalman-based motion model, finds where these anomalies correlate across independent vessels, and isolates the persistent regions into spoofing zones.}

We evaluate this framework on global AIS data from late November 2024 to early February 2025, \new{identifying} and clustering spoofing-like anomalies across more than 367,000 vessels worldwide. In total, the system identified over 17,000 \new{anomalous} episodes spanning 31,000 cumulative hours of \new{physically implausible} navigation. The resulting global map of 31 \new{ hotspots of anomalous GPS activity} reveals both expected patterns in conflict-adjacent waters, such as the Black Sea and Eastern Mediterranean, and surprising anomalies in civilian zones, like the Gulf of Mexico and Canary Islands.

The \emph{MSC Antonia} grounding underscores the operational significance of these findings. Our framework \new{identified} the same spoofing pattern in the Red Sea near Sudan, \new{at the identical location}, five months earlier in January 2025. \new{Similarly, we detected persistent spoofing in the Strait of Hormuz over a year before the 2026 Iran war drove widespread GPS interference across the Persian Gulf~\cite{iran_spoofing}}. This recurrence indicates that \new{both incidents were} part of a longer-lived spoofing zone our system could have identified in advance. Together, these results show that systematic measurement can explain known incidents and reveal emerging patterns of maritime GPS interference.

\noindent In summary, \textbf{our contributions} are as follows:
\begin{itemize}[leftmargin=*,nosep]
    \item The first global measurement of \new{large-scale} maritime GPS spoofing using real-world AIS data from over 367,000 vessels to quantify the prevalence, persistence, and geographic distribution of spoofing worldwide.
    \item A motion-aware, marine-specific \new{AIS anomaly analysis framework} that enables reliable identification of spoofing-like behavior in noisy AIS data by combining physically grounded trajectory modeling, cross-vessel consensus, and spatial clustering.
    \item An empirical characterization of 31 persistent \new{anomalous} hotspots, \new{of which at least 22 show strong evidence of GPS spoofing,} including regional case studies that reveal recurring spoofing patterns and their geopolitical context.
\end{itemize}

\section{Background and Related Work}
\label{sec:background}
This section summarizes the technical background of AIS and GPS spoofing and reviews prior work to contextualize our global measurement study.

\noindent\textbf{AIS and Maritime Cybersecurity.}
AIS is a transponder-based communication protocol mandated by the International Maritime Organization (IMO) for vessels exceeding 300 gross tons and all passenger ships~\cite{imo_ais}. Each AIS message broadcasts a vessel’s identity (including its Maritime Mobile Service Identity, or MMSI) together with GPS-derived position, speed, course, and voyage data~\cite{uscg_ais}. Because civilian GPS signals are typically unencrypted and unauthenticated, AIS-reported positions are inherently vulnerable to spoofing attacks~\cite{tippenhauer2011requirements,lee2019maturity}.

Prior work has shown that the maritime domain amplifies spoofing risks through insecure protocols and infrastructure, with demonstrated attacks against AIS~\cite{balduzzi2014security}, satellite terminals~\cite{pavur2020tale}, marine communication stacks~\cite{tran2021marine}, and cyber-physical ship control systems~\cite{progoulakis2021cyber}. Broader analyses emphasize the historically underexplored state of maritime cybersecurity~\cite{direnzo2015little}. More recently, several studies have leveraged AIS data to detect spoofing or falsified vessel behavior, including protocol-level validation, trajectory-based anomaly detection, and forensic or domain-specific analyses~\cite{louart2023detection,zheng2023identification,androjna2021ais,androjna2023ais}. While these efforts demonstrate that AIS can reveal spoofing, they remain localized or attack-specific, and do not provide a global measurement of spoofing prevalence or persistence. Our work fills this gap by systematically measuring maritime GPS spoofing at global scale using real-world AIS traffic.

\noindent\textbf{GPS Spoofing and Detection Techniques.}
GPS\footnote{Although the broader term GNSS encompasses multiple constellations, we use ``GPS'' generically since AIS does not specify the system.} spoofing involves transmitting counterfeit satellite signals to manipulate a receiver’s perceived position or time~\cite{bhatti2017hostile,xu2023sok,xu2024physcout}. Unlike jamming, spoofing preserves apparent functionality while providing false coordinates~\cite{clover_russia_gps_jamming,gahnstrom_jamming_spoofing}. In maritime settings, spoofing can induce kilometer-scale displacements or false convergence that violate physical motion constraints~\cite{spravil2023detecting,jones_black_sea_spoofing,zorri_pnt_weaponization}, in contrast to multipath interference, which typically produces meter-scale deviations~\cite{weill1997conquering,esa_navipedia_multipath}.

Extensive research has demonstrated GPS spoofing across automotive, aerial, and autonomous domains~\cite{tippenhauer2011requirements,psiaki2016gnss,zeng2018all,sathaye2022experimental,narain2019security,shen2020drift,yang2023location,tibaldo2025gnss,zhang2025ghost}. A wide range of detection and mitigation techniques have been proposed, including angle-of-arrival filtering~\cite{liu2021stars}, vision-based cross-checks~\cite{davidovich2022visas}, statistical anomaly detection~\cite{amro2022navigation}, and specialized anti-spoofing receivers~\cite{sathaye2022semperfi}. Crowd-GPS-Sec detects attacks using crowdsourced aircraft data~\cite{jansen2018crowd}, but requires dense multilateration coverage unavailable at sea. Complementary approaches include secure LEO-based ranging~\cite{coppola2025leo} and distributed spoofers that enhance stealth and range~\cite{cheng2025distributed}.

Collectively, these studies establish that GPS spoofing is technically feasible and operationally disruptive, yet none quantify its global prevalence or persistence in the maritime domain. Our work provides the first systematic, global measurement of maritime GPS spoofing by analyzing motion anomalies across worldwide AIS traffic and aggregating them into persistent spoofing hotspots.

\section{Methodology}
\new{We first define our threat model (Section~\ref{sec:threatmodel}) and datasets (Section~\ref{sec:dataset}), then describe our pipeline whose two stages mirror the invariants of regional external spoofing introduced in Section~\ref{sec:intro}. Stage 1 identifies the per-vessel invariant: using a Kalman-based motion model, we flag physically implausible motion in each vessel's trajectory (Section~\ref{sec:part1}). Stage 2 identifies the cross-vessel and spatial invariants: we cluster these anomalies by spatial and temporal overlap into persistent zones, separating interference that affects many vessels at once from irregularities that affect only one (Section~\ref{sec:part2}). We then discuss the limitations of this design (Section~\ref{sec:limitations}) and validate the pipeline across global maritime traffic (Section~\ref{sec:validation}).}

\subsection{Threat Model}
\label{sec:threatmodel} 
\begin{figure}[]
    \centering
    \includegraphics[width=0.40\textwidth]{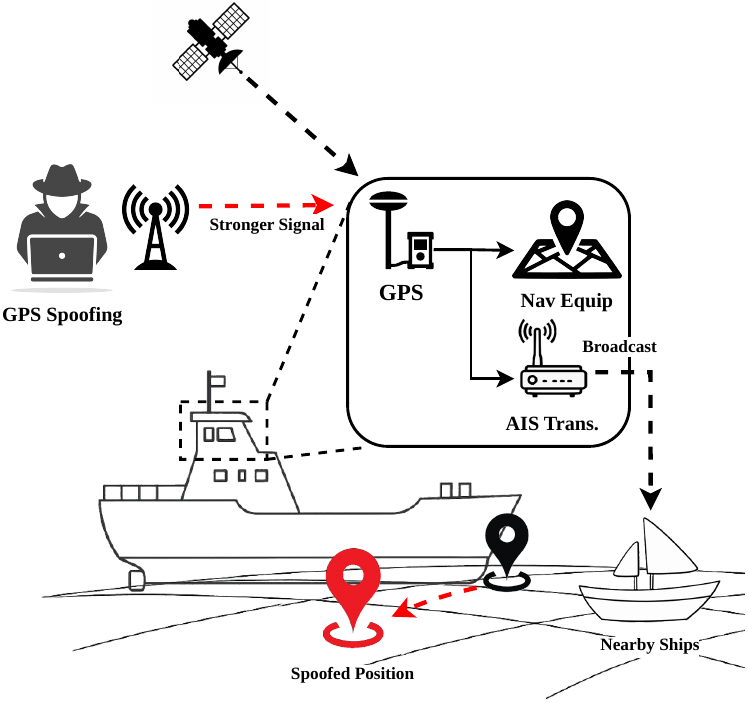}
    \caption{\new{Maritime GPS spoofing threat model. A nearby terrestrial or airborne transmitter overrides genuine GPS signals with a stronger counterfeit, causing the vessel's navigation system to compute a false position that the AIS transponder broadcasts to nearby ships.}}
    \label{fig:threat_model}
\end{figure}

\new{Our threat model is large-scale external GPS spoofing: many vessels' reported positions are displaced at once, rather than any one ship. The adversary broadcasts a counterfeit GPS-like signal stronger than the authentic satellite transmissions at the receiver, causing a vessel's navigation system to compute an erroneous position; the navigation system forwards this false fix to the AIS transponder, which broadcasts the spoofed coordinates to nearby ships and monitoring services (Figure~\ref{fig:threat_model}). Because the source acts over a bounded area, it affects many vessels within its footprint simultaneously, inducing the kilometer-scale, regionally coherent displacements our measurement targets. This threat model is grounded in documented real-world operations: large-area spoofing affecting many vessels simultaneously has been reported in the Black Sea~\cite{jones_black_sea_spoofing,c4ads_gps_spoofing_russia_syria}, off the Chinese coast~\cite{harris_gps_mystery,trevithick_gps_spoofing_china}, and recently in the Strait of Hormuz~\cite{iran_spoofing}.}

\new{Outside this threat model are anomalies confined to a single vessel, whether from faulty transponders, recycled MMSI identifiers, or deliberate falsification of AIS messages while the onboard GPS computes correctly, which we term \emph{self-spoofing}. Our pipeline separates these from external interference, since each affects one vessel rather than many across a region (Section~\ref{sec:part2}); we examine self-spoofing further in Section~\ref{sec:case_studies}.}

\subsection{Dataset}
\label{sec:dataset}

\new{Our study draws on three AIS datasets that share the same record structure, each providing per-vessel timestamps, position, speed, course, heading, identity, and operational status. Because no public AIS feed offers global coverage, we center our measurement on a global commercial feed from Spire Global and draw on two public regional feeds, from NOAA and the Danish Maritime Authority, that span U.S. and European waters~\cite{noaa_data,dma_data}.

Spire collects AIS data from low-Earth orbit satellites, providing near-real-time global coverage used operationally by national authorities and governments~\cite{spire_global,spire_ca_contract_2024}. Its feed spans late November 2024 to early February 2025 and integrates both satellite and ground-based reception, giving comprehensive global coverage and precise timestamp alignment with minimal reporting gaps. Over this period it comprises 379,416 unique vessels, of which 125,569 appeared in both December and January. Importantly, the data captures where vessels report themselves to be rather than their independently verified locations, making it well suited to analyzing positional inconsistencies and identifying spoofing-consistent behavior.

The two public feeds are regional but comparable to Spire in format and fidelity. Running our pipeline on them reproduces our findings in both the U.S. and Europe from open data alone, confirming that the results do not depend on any single source and supporting independent verification.}

\subsection{\new{Stage 1: Per-Vessel Anomaly} Analysis}
\label{sec:part1}

\new{This stage analyzes each vessel in isolation. We treat each vessel identifier (MMSI) as a single vessel and flag motion that violates physical or environmental plausibility, such as abrupt jumps, implausible speeds, or movement over land. At this stage, we do not attribute a cause: a flagged trajectory may reflect external interference, deliberate self-spoofing, or a faulty transponder. Stage 2 (Section~\ref{sec:part2}) distinguishes these by testing whether the same anomaly appears across many vessels at once.}

\begin{figure}[]
    \centering
    \includegraphics[width=0.47\textwidth]{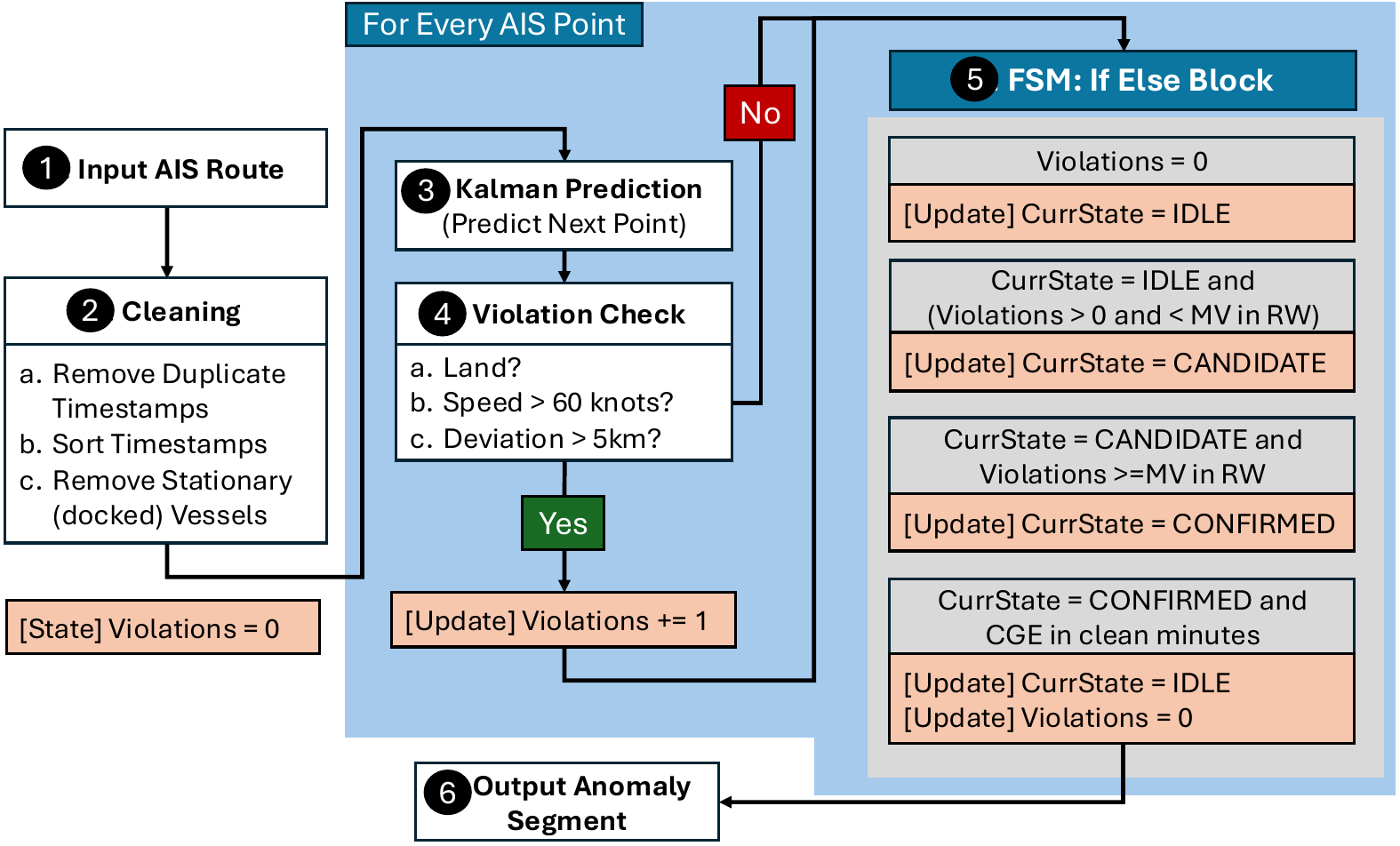}
    \caption{Overview of the per-ship anomaly-\new{identification} stage}
    \footnotesize
    MV = Minimum violations; RW = Rolling window; CGE = Clean gap end
    \label{fig:method_part1}
\end{figure}

As shown in Figure~\ref{fig:method_part1}, the per-ship anomaly-\new{identification} pipeline proceeds through six steps: (\blackcircled{1}) input AIS parsing, (\blackcircled{2}) cleaning, (\blackcircled{3}) Kalman prediction, (\blackcircled{4}) violation checks, (\blackcircled{5}) finite-state grouping, and (\blackcircled{6}) output generation. These stages transform raw AIS tracks into temporally coherent \emph{\new{anomaly} episodes} characterized by physically implausible motion or location patterns. The following subsections describe each component in detail.

\noindent\textbf{\blackcircled{1}-\blackcircled{2} Input AIS Route and Cleaning.} Each vessel’s raw AIS data is parsed into timestamp-ordered trajectories containing latitude, longitude, speed, and course fields. 
We sort entries by time and deduplicate reports that share the same timestamp and position. Finally, vessels that remain stationary within a 0.5\,km radius are discarded, filtering out AIS base stations and permanently moored transmitters while tolerating minor GPS jitter.

\noindent\textbf{\blackcircled{3} Kalman Prediction.} 
To estimate intermediate vessel positions and suppress measurement noise, we employ a linear Kalman filter\footnote{Kalman filters are widely used in navigation and sensor fusion, including satellite positioning and real-time tracking~\cite{wendel2001direct,alabsi2021tracking,nishad2025advanced}.} for trajectory state estimation. The filter alternates between a \emph{prediction} step (propagating a motion model forward) and an \emph{update} step (correcting with new AIS observations), optimally weighting model dynamics and sensor reliability under Gaussian noise assumptions~\cite{levy1997kalman}. \cam{Given the reporting cadence of AIS transmissions, which ranges from 2 seconds to 3 minutes depending on vessel class and speed~\cite{itu_M.1371-5},} we adopt a constant-velocity motion model that assumes each vessel maintains its speed and heading between updates. Because ships exhibit high inertia and rarely undergo abrupt acceleration or deceleration, this assumption provides a realistic short-term approximation of vessel motion. Process and measurement noise covariances were set to $10^{-5}$ and $10^{-4}$, respectively, corresponding to roughly kilometer-scale GPS uncertainty, so the filter tracks smooth motion yet remains responsive to course changes.

\noindent In the Kalman formulation, the vessel’s state vector 
$\mathbf{x}_t=\begin{bmatrix}\phi_t & \lambda_t & \dot{\phi}_t & \dot{\lambda}_t\end{bmatrix}^{\!\top}$ 
encodes latitude ($\phi_t$), longitude ($\lambda_t$), and their time derivatives (velocity components). 
The measurement vector 
$\mathbf{z}_t=\begin{bmatrix}\phi_t & \lambda_t\end{bmatrix}^{\!\top}$ 
contains the observed AIS positions. 
During the \emph{prediction} step, the state-transition matrix $\mathbf{F}(\Delta t)$ advances the previous position forward in time using $\Delta t$ (in hours), while process noise $\mathbf{w}_t$ models unobserved accelerations. 
In the \emph{update} step, the measurement matrix $\mathbf{H}$ projects the predicted state to observation space and corrects it with the new AIS reading, subject to measurement noise $\mathbf{v}_t$. 

\noindent\textbf{Model equations.} The resulting linear system is:
\begin{equation}
\mathbf{x}_{t+\Delta t}= \mathbf{F}(\Delta t)\,\mathbf{x}_t+\mathbf{w}_t,\qquad
\mathbf{z}_t=\mathbf{H}\,\mathbf{x}_t+\mathbf{v}_t,
\end{equation}
\begin{equation}
\mathbf{F}(\Delta t)=
\begin{bmatrix}
1&0&\Delta t&0\\
0&1&0&\Delta t\\
0&0&1&0\\
0&0&0&1
\end{bmatrix},\qquad
\mathbf{H}=
\begin{bmatrix}
1&0&0&0\\
0&1&0&0
\end{bmatrix},
\end{equation}
with $\mathbf{w}_t\!\sim\!\mathcal{N}(0,\mathbf{Q})$ and $\mathbf{v}_t\!\sim\!\mathcal{N}(0,\mathbf{R})$. Longitude differences are normalized to the principal branch to handle $\pm180^\circ$ wrap. 

\noindent At each update, the predicted position $(\hat{\phi}_t^{-},\hat{\lambda}_t^{-})$ is compared with the new AIS observation $(\phi_t,\lambda_t)$ to compute a geodesic residual
$e_t \;=\; d_{\text{hav}}\!\bigl(\hat{\phi}_t^{-},\hat{\lambda}_t^{-};\,\phi_t,\lambda_t\bigr),$
where $d_{\text{hav}}$ is the Haversine great-circle distance between two latitude-longitude coordinates:

\begin{equation}
    \label{eq:haversine}
    \begin{aligned}
    &a \coloneqq \sin^2\!\frac{\phi_2-\phi_1}{2}
       + \cos\phi_1\,\cos\phi_2\,\sin^2\!\frac{\lambda_2-\lambda_1}{2},\\
    &d_{\text{hav}}(\phi_1,\lambda_1;\phi_2,\lambda_2)
    = 2 R_\oplus \arcsin\!\bigl(\sqrt{a}\,\bigr)
    \end{aligned}
\end{equation}

\noindent where $R_\oplus = 6371$\,km is the Earth's mean radius. 
This formulation avoids distortions introduced by planar projections and ensures that deviation thresholds (e.g., 5\,km) correspond to true geodesic distances globally~\cite{dbski_haversine,bansal2021deriving}. 
A \emph{deviation} violation is triggered when $e_t > 5\,\text{km}$, indicating that the vessel’s reported position departs significantly from its predicted path.

Because estimation error accumulates rapidly with sparse updates, the filter is re-initialized whenever two consecutive AIS messages are separated by more than seven minutes. 
As shown in Figure~\ref{fig:time_diff} in the appendix, over 92\% of timestamp gaps in our dataset are shorter than this threshold, making seven minutes a natural empirical cutoff that balances continuity with stability.

\noindent\textbf{\blackcircled{4} Violation Checks.} Each cleaned trajectory is analyzed at minute-level granularity using three \new{violation checks}: (1) \emph{Land points}, locations outside a coastal-water buffer using the Global Surface Water Occurrence dataset with a 1\,km buffer to avoid overflagging near coastlines~\cite{pekel2016high,jrc_dataset}; (2) \emph{Deviation points}, positions whose Kalman-predicted and reported locations differ by more than 5\,km, indicating motion inconsistent with realistic vessel dynamics; and (3) \emph{Speed violations}, instantaneous velocities above 60\,knots (kn).

We enforce physical feasibility constraints using empirically validated thresholds. For speed, typical large commercial vessels operate well below 30\,kn in service, with design speeds around 19-25\,kn for common container-ship classes~\cite{cedelft_slow_steaming_2012}, and fast patrol craft commonly reach sprint speeds in the 35-45\,kn range~\cite{uscg_rbm,usn_mkvi,shaldag_mk2}; we therefore conservatively set 60\,kn as an upper bound for physically plausible movement. For the deviation check, the 5\,km cutoff was derived from the dataset-wide distribution of Kalman residuals (Figure~\ref{fig:error_data} in the appendix). Before filtering, 98.5\% of residuals fall below 5\,km, yet 36.1\% of vessels exhibit at least one extreme outlier above this limit (5-10\,km: 18.4\%; 10-20\,km: 5.9\%; 20-50\,km: 2.6\%; $>$50\,km: 9.2\%). After removing physically implausible points, those with on-land positions or speeds exceeding 60\,kn, the remaining clean dataset shows only 0.3\% of residuals above 5\,km; thus, 5\,km separates normal positional noise from rare, large deviations that persist even after conservative filtering. We deliberately avoid a smaller cutoff (e.g., 1-2\,km) because benign GPS and AIS timing errors, coastal multipath reflections, reporting jitter, and ionospheric disturbances during geomagnetic storms can reach hundreds of meters~\cite{weill1997conquering,esa_navipedia_multipath,noaa_weather}; a tighter threshold would increase false positives without improving sensitivity to spoofing.

\noindent\textbf{\blackcircled{5} Temporal Grouping via Finite-State Machine (FSM).} To aggregate point-level anomalies into temporally coherent \new{anomaly} episodes, we employ an FSM that enforces persistence and recovery rules. The FSM operates on minute-binned AIS data and formalizes episode \new{identification} using three parameters: Minimum Violations ($MV$), a Rolling Window ($RW$), and a Clean Gap End ($CGE$); see Figure~\ref{fig:method_part1}.

Each minute is labeled anomalous if any of its AIS points trigger a land, speed, or deviation violation.  
We then slide a 30-minute rolling window ($RW$) across time and compute the ratio $r_t = A_{\text{RW}} / N_{\text{RW}}$, where $A_{\text{RW}}$ is the number of anomalous points and $N_{\text{RW}}$ is the total points in that window.  
A minimum violation threshold ($MV$) is defined as
$\mathrm{MV} = \lceil 0.70 \times N_{\text{RW}} \rceil$, 
requiring at least $70\%$ of points in a full 30-minute window to be anomalous before confirming an episode. This conservative cutoff was selected from the dataset-wide window-ratio distribution, where fewer than 10\% of all 30-minute windows exceed the 70\% mark. As shown in Figure~\ref{fig:window_size} in the appendix, the curve flattens well below this level, making 70\% a deliberately strict threshold that minimizes false positives while still capturing clear, persistent spoofing.

\noindent\textbf{State Logic.}
The FSM progresses through three phases that capture the temporal structure of spoofing activity:

\begin{itemize}[leftmargin=1.5em,nosep]
    \item \textbf{IDLE:} The default monitoring phase in which no sustained anomalies are present. For each new minute, the rolling window is updated. If a few anomalies appear ($0 < A_{\text{RW}} < \mathrm{MV}$), the FSM transitions into the \texttt{CANDIDATE} state to monitor whether the anomalies persist or not.
    \item \textbf{CANDIDATE:} A provisional phase that is triggered when anomalies begin to accumulate but remain below the threshold. If the quorum condition ($r_t \ge 0.70$ in a fully covered 30-minute window) is reached, the FSM escalates to an \texttt{ACTIVE} state; otherwise, if anomalies subside before this threshold is met, it returns to \texttt{IDLE}.
    \item \textbf{ACTIVE:} A confirmed \new{anomaly} episode, which persists until a clean streak of $120$ consecutive minutes ($CGE$) with no anomalies occurs, marking recovery and returning the FSM to \texttt{IDLE}.
\end{itemize}

This temporal logic encoded in the FSM ensures that transient spikes or isolated errors do not trigger false positives. Short-lived or low-density anomaly bursts remain contained within the \texttt{CANDIDATE} phase and naturally expire, whereas \new{persistent, high-density anomalies} satisfy both the $MV$ ($\ge 70\%$ in $30$\,min) and $CGE$ ($120$ clean\,min) criteria before being confirmed. The 120-minute clean-gap threshold was chosen empirically: 61.5\% of inter-event gaps occur within two hours, after which the distribution drops sharply (Figure~\ref{fig:clean_gap} in the appendix). This cutoff provides a conservative recovery window that separates consecutive \new{anomaly} events without merging unrelated ones. The Start-by-Quorum / End-by-Consecutive formulation provides temporal stability in high-noise AIS environments. \cam{The four parameters governing this stage are derived from the observed data distributions; we test their influence on our findings in Section~\ref{sec:validation}.}

\noindent\textbf{\blackcircled{6} Outputs.} Each \texttt{ACTIVE} period yields a single episode record with start and end timestamps, duration, anomaly-type histograms, and geometric deviations between observed and Kalman-predicted paths. Episodes shorter than 30 minutes or containing fewer than one AIS sample are discarded. The resulting set of per-vessel \new{anomaly} segments forms the baseline input for the subsequent cluster-level analysis.

\subsection{\new{Stage 2: Anomaly}-Zone Clustering}
\label{sec:part2}

\begin{figure}[]
    \centering
    \includegraphics[width=0.48\textwidth]{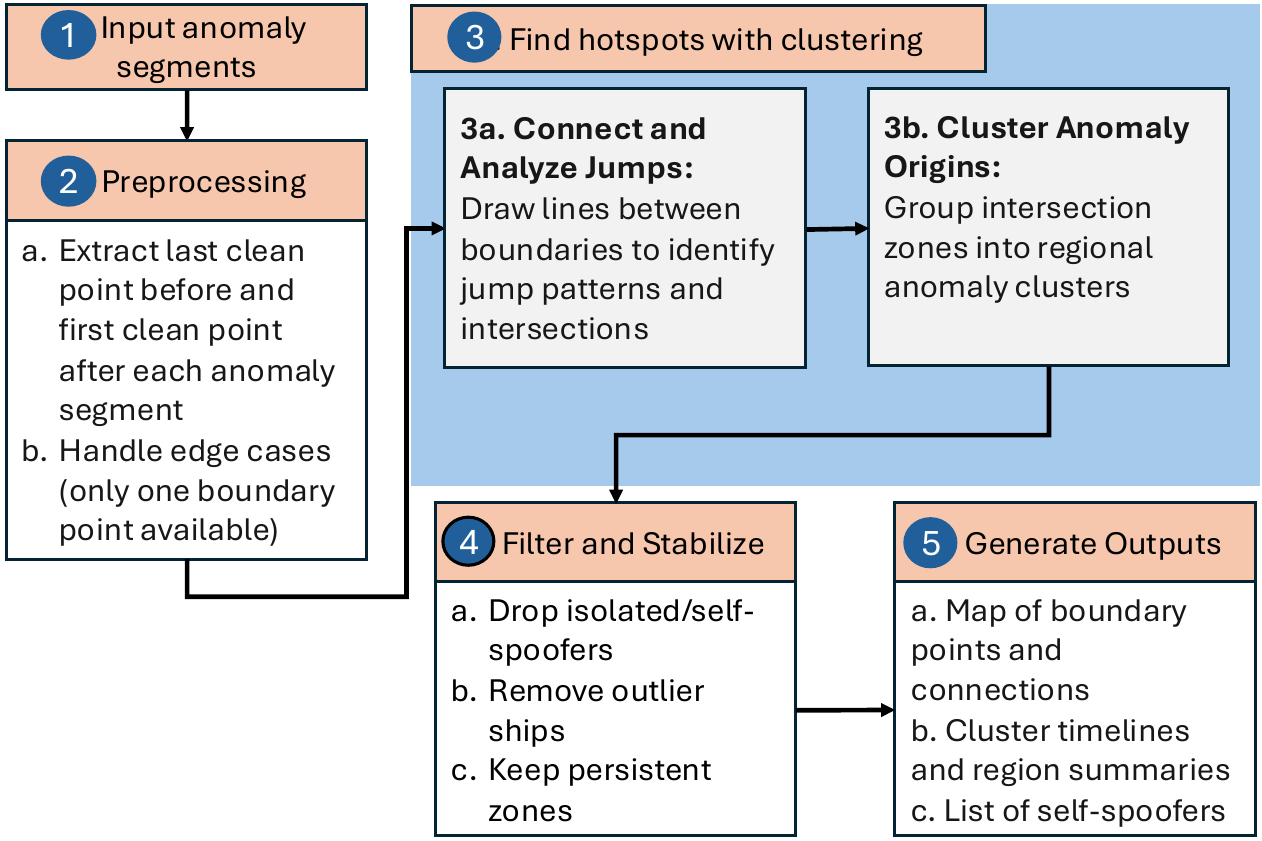}
    \caption{Overview of \new{anomaly}-zone clustering/attribution stage.}
    \label{fig:method_part2}
\end{figure}

This stage aggregates \new{anomaly} segments from Stage 1 to \new{identify regional patterns} and probable interference origins. It transitions from vessel-level analysis to multi-vessel inference, combining spatial geometry, temporal overlap, and density-based clustering to reveal persistent \new{anomaly} zones.

\noindent\textbf{\bluecircled{1} Input \new{Anomaly} Segments.}  
All \new{anomaly} segments extracted in Stage 1 serve as inputs to this stage. Each segment captures a temporally bounded interval of physically implausible motion, including start and end timestamps, vessel identifiers, and segment geometry.

\noindent\textbf{\bluecircled{2} Preprocessing.}  
For every \new{anomaly} segment, we extract the last clean AIS point preceding the episode and the first clean point following it. These boundary points delineate the apparent entry and exit of \new{anomalous} motion, effectively the ``jump endpoints'' of the \new{anomalous movement}. When only one boundary is available (e.g., truncated or missing trajectories), it is retained as a single-ended edge to preserve partial spatial information.

\noindent\textbf{\bluecircled{3} Jump-Line Construction and Clustering.}  
This step connects vessel-level \new{anomaly} episodes into shared regional patterns using a two-part process. First, each episode’s boundary points are connected to form a ``jump line’’ representing the vessel’s apparent displacement during \new{the anomaly episode}. Spatial intersections among multiple jump lines mark candidate regions where independent vessels exhibit correlated \new{anomalous} behavior, an indicator of common false coordinates. This helps us connect per-vessel \new{anomaly} events to group-level patterns.

Next, we apply DBSCAN to cluster nearby intersections into potential \new{anomaly} zones~\cite{ester1996density,schubert2017dbscan}. This non-parametric algorithm groups points that lie within a fixed spatial neighborhood of each other, forming clusters when at least a minimum number of nearby intersections (\textit{MinPts}) occur within a radius $\varepsilon$. Here, each point corresponds to the intersection of jump lines (latitude-longitude coordinates). \cam{We set $\varepsilon = 50$\,km and \textit{MinPts} $= 5$.

The relevant spatial scale for $\varepsilon$ is the footprint of an interfering emitter. Ground-based reach is bounded by the radio horizon: with the effective Earth radius factor $k = 4/3$~\cite{itu834} and a mean Earth radius of 6371\,km, the line-of-sight distance is $d \approx 4.12(\sqrt{h_t} + \sqrt{h_r})$\,km with antenna heights in metres. A transmitter 10 to 100\,m above ground and a vessel GPS antenna 10 to 30\,m above the waterline give 26 to 64\,km, placing 50\,km mid-range. Airborne emitters have wider footprints, so this choice fragments a single footprint across clusters instead of merging distinct sources. For \textit{MinPts}, prior work~\cite{ester1996density,schubert2017dbscan} reports 4 as a standard default for two-dimensional data, and larger values are recommended for very large or noisy datasets~\cite{schubert2017dbscan}; we set 5 on that basis.

We do not rely on these arguments alone. We swept $\varepsilon$ from 25\,km, below the shortest horizon above, to 100\,km, twice our value, and \textit{MinPts} from 3, below the published default, to 10, twice ours (Table~\ref{tab:dbscan-sensitivity}). All 17 geographic regions are recovered for $\varepsilon$ between 40 and 75\,km and for \textit{MinPts} between 3 and 7. Outside that window clusters merge at 100\,km and sparse zones no longer reach the density requirement at 25\,km. Repeating the measurement with many-to-one matching, which permits one cluster to cover several zones, changes the counts by at most three. Our results are therefore not sensitive to the choice of $\varepsilon$ or \textit{MinPts} within roughly a factor of two of the chosen values.}

\noindent\textbf{\bluecircled{4} Filter and Stabilize.}  
To retain only the meaningful and persistent \new{anomaly} zones identified in the previous step, we apply three stabilization filters. First, vessels that never share spatial or temporal overlap with others are excluded as \textit{self-spoofers} or individual reporting faults.
Second, we remove spatial outliers whose \new{anomalous} locations lie more than three times the mean intra-cluster distance from the centroid, a standard outlier-rejection heuristic that filters detached points without fragmenting dense clusters. On average, clusters contained only 1\% of such outliers, arising when jump lines from unrelated vessels overlapped in the same region, causing DBSCAN to group them despite representing distinct \new{anomaly} incidents.
Finally, clusters that recur consistently across multiple time windows are merged and marked as \textit{persistent zones}, indicating stable or repeated \new{anomaly} activity in that region.

\noindent\textbf{\bluecircled{5} Generate Outputs and Attribution Labels.} Building on the filtered \new{anomaly} zones from the previous step, we assign vessel-level classification labels to distinguish coordinated \new{anomalies} from isolated or faulty transmissions. Each vessel receives one of two primary classification labels. Vessels whose anomalous segments coincide spatially and temporally with others inside a persistent cluster are labeled \new{\emph{cluster-associated}}, reflecting \new{potential} coordinated or externally induced spoofing. Episodes occurring alone, outside any cluster, are labeled \emph{self-spoofed or faulty}.

We further subdivide the self-spoofed/faulty category to distinguish between likely \textit{MMSI reusers} and genuinely malfunctioning transmitters. Specifically, vessels that retain the same MMSI but change associated metadata such as name, callsign, or flag are classified as probable MMSI reusers, suggesting recycled identifiers across different ships. In contrast, vessels with stable identifiers yet unusually high flag rates are labeled as faulty transmitters, indicative of persistent AIS hardware or reporting errors. The final outputs include maps linking \new{anomaly}-segment boundaries, temporal cluster timelines, and a list of self-spoofers.

\subsection{Limitations}
\label{sec:limitations}
\new{Our framework targets large-scale regional spoofing and may miss stealthy interference that evolves within plausible bounds. \cam{Concretely, four classes of activity fall outside detection by construction: displacements smaller than the 5\,km deviation threshold, episodes shorter than the 30\,min candidate window, windows in which fewer than 70\% of points are anomalous, and interference affecting only a single vessel, which Stage~2 excludes as self-spoofing or a reporting fault. Each of these bounds removes activity rather than adding it.} Our measurements therefore represent a lower bound on global spoofing activity. Our time window also cannot capture seasonal or longer-term trends. \cam{AIS reports the position a vessel's receiver computed, not the signal it received, so we cannot localize transmitters, attribute interference to specific actors, or observe the spoofing-system implementation and signal-generation mechanism. Our geopolitical context and attacker objectives (Section~\ref{sec:case_studies}), such as air-defense spillover near Gaza, are therefore inferred as hypotheses, and confirming them requires RF measurement or operator accounts.}}

\subsection{Validation}
\label{sec:validation}
\new{Because no onboard ground truth exists for maritime GPS spoofing at scale, we validate our findings through complementary checks: ruling out alternative causes of AIS anomalies, \cam{testing the sensitivity of our results to the detection parameters,} testing for over-flagging in dense traffic, distinguishing external interference from self-spoofing, and corroborating identified zones against documented incidents.

\begin{table}[]
\centering
\caption{Alternative explanations excluded at each stage of the path a position report travels, from satellite signal to aggregated dataset. These checks apply uniformly across all 31 anomaly zones before tier assignment.}
\label{tab:alternatives-excluded}
\footnotesize
\setlength{\tabcolsep}{4pt}
\renewcommand{\arraystretch}{1.05}
\begin{tabular}{@{}p{3.0cm} p{4.6cm}@{}}
\toprule
\textbf{Alternative} & \textbf{Pipeline Exclusion Mechanism} \\
\midrule
\multicolumn{2}{@{}l}{\underline{\textbf{\textsc{signal path}}}} \\
\addlinespace[1pt]
Space-weather effects & Tens-of-meter errors, far below 5\,km threshold (Section~\ref{sec:part1}) \\
GPS jamming & Causes loss of fix and reporting gaps, not coherent false tracks (Section~\ref{sec:background}) \\
\addlinespace[3pt]
\multicolumn{2}{@{}l}{\underline{\textbf{\textsc{gps receiver}}}} \\
\addlinespace[1pt]
Multipath / jitter & Hundreds-of-meter errors, below 5\,km threshold (Section~\ref{sec:part1}) \\
Isolated receiver errors & FSM persistence: 70\% anomalous over 30\,min + 120\,min recovery (Section~\ref{sec:part1}) \\
Persistent receiver errors & Affects one vessel only, excluded by cross-vessel clustering (Section~\ref{sec:part2}) \\
\addlinespace[3pt]
\multicolumn{2}{@{}l}{\underline{\textbf{\textsc{ais transponder}}}} \\
\addlinespace[1pt]
Faulty transmitters & High anomaly rate with consistent metadata (Section~\ref{sec:part2}) \\
MMSI reuse & Metadata consistency classification (Section~\ref{sec:part2}) \\
Single-vessel \mbox{self-spoofing} & Affects one vessel only, excluded by cross-vessel clustering (Section~\ref{sec:part2}) \\
\addlinespace[3pt]
\multicolumn{2}{@{}l}{\underline{\textbf{\textsc{aggregate data}}}} \\
\addlinespace[1pt]
Coordinated self-spoofing & Vessel diversity: median 30 flag states, 5 ship types per cluster (Table~\ref{tab:cluster_diversity}) \\
Over-flagging in dense traffic & Validation across 22 ports with no reported spoofing (Table~\ref{tab:port_validation}) \\
\bottomrule
\end{tabular}
\end{table}

\noindent\textbf{Filtering Out Alternative Root Causes.}
Several phenomena besides external spoofing can produce anomalous AIS positions. Each must arise somewhere along the path a position report travels: satellite signals propagate to the vessel, the GPS receiver computes a fix from them, the AIS transponder broadcasts that fix under a vessel identity, and our data sources aggregate the reports. Our pipeline excludes the alternatives at every stage. Along the signal path, ionospheric disturbances during even severe geomagnetic storms degrade GPS accuracy on the order of tens of meters~\cite{noaa_weather}, and multipath and jitter at the receiver cause deviations of at most a few hundred meters~\cite{weill1997conquering,esa_navipedia_multipath}; both fall two orders of magnitude below our 5\,km deviation threshold. Jamming, the other deliberate form of signal interference, produces loss of fix and reporting gaps rather than coherent false tracks (Section~\ref{sec:background}). Isolated receiver errors are removed by the FSM persistence requirement (Section~\ref{sec:part1}). Persistent receiver errors affect only a single vessel and are excluded by the cross-vessel clustering requirement, alongside self-spoofing (Section~\ref{sec:part2}). At the AIS transponder, faulty hardware and reused MMSI identities are separated by metadata-consistency checks (Section~\ref{sec:part2}). At the aggregation stage, over-flagging in dense traffic and coordinated self-spoofing across many vessels are tested directly later in this section. Table~\ref{tab:alternatives-excluded} summarizes each alternative and their exclusion.}

\noindent\textbf{Validation in Dense Maritime Traffic.}
To evaluate robustness under dense vessel activity and potential GPS multipath interference, we analyzed AIS traffic in 22 of the world’s busiest commercial ports (Table~\ref{tab:port_validation} in the appendix)~\cite{wsc_top50,lloyds_top100}. Across all ports, none of which \cam{have reported spoofing}, the median fraction of vessels flagged \new{by our pipeline} is below 0.1\%, and no port exceeds 1\%. High-traffic hubs such as Los Angeles, Rotterdam, and Singapore exhibit similarly low rates, indicating that vessels operating in dense maritime environments are not systematically misclassified. We further find that the small number of residual false positives are primarily attributable to AIS artifacts such as MMSI reuse or faulty transmitters, rather than algorithmic errors (see Appendix~\ref{sec:appendix_validation} for more analysis).

\noindent\textbf{External vs. Self-spoofing Validation.}
To assess whether \new{identified} clusters could plausibly arise from coordinated self-spoofing, we examine vessel diversity and pattern consistency within each cluster. Across clusters, \new{flagged} vessels span a wide range of flag states and ship types, with a median of 30 distinct flag states and 5 vessel classes per cluster, making coordinated self-spoofing unlikely given the need for independent operators across jurisdictions and vessel types to deploy similar behavior simultaneously (Table~\ref{tab:cluster_diversity}). Moreover, vessels within each cluster exhibit a shared spoofing\new{-like} geometry, with a median of 72\% of vessels per cluster following the same displacement pattern despite this operational diversity. We note that one cluster is dominated by a single flag state, which is expected given its location in inland China; however, this cluster still spans multiple vessel types and has consistent spoofing\new{-like} geometry, supporting external-interference instead of coordinated self-spoofing.

\cam{\noindent\textbf{Parameter Sensitivity.} Four Stage-1 parameters are derived from the observed data distributions (Section~\ref{sec:part1}). To test whether our findings depend on those choices, we ran a parameter sweep, varying each independently over a wide range. Recovery is measured geographically. A zone counts as recovered if a cluster from the swept setting falls within 100\,km of its baseline center, twice the clustering radius, with each cluster matched to at most one zone. The criterion is insensitive to whether a zone subdivides or merges. Some zones also lie close enough that one cluster may cover several, so we additionally group the 31 zones into the 17 geographic regions they occupy, such as the Black Sea or the Gulf of Mexico, and report how many regions are recovered. The deviation threshold, the parameter most directly tied to our detection criterion, has no effect: all 31 zones are recovered at every value from 2 to 10\,km, even though the underlying episode count varies by a third across that range. Across the 20 parameter configurations tested, all 17 geographic regions are recovered in 13 of those configurations, with the full 30-360\,min CGE range and the 50-70\% MV range leaving the identified regions unchanged. Losses are confined to the smallest zones: outside the two most aggressive settings (MV\,$=90\%$, reset\,$=3$\,min), every zone that drops out has 12 or fewer vessels, while every zone with more than 12 vessels is recovered in every configuration. Detection therefore degrades as vessel traffic decreases. Appendix~\ref{sec:appendix_validation} reports the full parameter sweep.}

\new{\noindent\textbf{Corroboration Against Documented Incidents.}
Finally, we cross-reference identified zones against independent incident reports and conflict timelines. As detailed in Section~\ref{sec:evidence_grading}, 13 of 31 zones align with externally documented spoofing, providing ground-truth-adjacent confirmation that the pipeline recovers real interference.}

\section{Results}

Below, Section~\ref{sec:detection_and_characterization} quantifies the global scope and structure of \new{anomaly activity and grades the spoofing inference for each zone}, while Section~\ref{sec:case_studies} interprets these patterns through regional case studies that connect \new{identified} hotspots to real-world events, including conflict-linked interference along the Gaza coast and the Red Sea pattern preceding the \emph{MSC Antonia} grounding.

\subsection{\new{Identification} and Characterization}
\label{sec:detection_and_characterization}

Applying our vessel-level anomaly \new{analysis} (Section~\ref{sec:part1}) to the late November 2024-early February 2025 global AIS dataset produced 17,936 \new{anomaly} episodes across 2,663 distinct vessels, totaling 31,328 hours of anomalous operation. Most episodes arose from deviation (52\%) or speed (45\%) violations, while only 3\% involved physically impossible land positions. Deviation anomalies correspond to vessels whose reported positions deviated by more than 5 km from their Kalman-predicted trajectories, whereas speed anomalies reflect instances where instantaneous velocity exceeded 60~knots, surpassing the physically plausible range for vessels. Recall that we define an \new{anomaly} episode as any interval in which at least 70\% of reported positions within a 30-minute window violate one or more anomaly constraints, a conservative criterion later used for clustering.

\begin{table}[]
    \centering
    \caption{Distribution of \new{anomaly} episode durations (ED*) and number of \new{anomaly} episodes per vessel (EpV$^\diamond$).}
    \label{tab:duration_exposure}
    \begin{minipage}{0.5\linewidth}
    \centering
    \begin{tabular}{lrr}
    \toprule
    \textbf{ED*} & \textbf{Count} & \textbf{\%} \\
    \midrule
    $<$60 min & 12{,}702 & 70.8 \\
    1–2 h & 2{,}468 & 13.8 \\
    2–4 h & 1{,}372 & 7.6 \\
    4–8 h & 812 & 4.5 \\
    8–24 h & 514 & 2.9 \\
    $>$24 h & 68 & 0.4 \\
    \bottomrule
    \end{tabular}
    \end{minipage}\hfill
    \begin{minipage}{0.5\linewidth}
    \centering
    \begin{tabular}{lrr}
    \toprule
    \textbf{EpV$^\diamond$} & \textbf{Count} & \textbf{\%} \\
    \midrule
    1 & 1{,}439 & 54.0 \\
    2 & 303 & 11.4 \\
    3–4 & 249 & 9.4 \\
    5–9 & 261 & 9.8 \\
    10–19 & 198 & 7.4 \\
    $\ge$20 & 213 & 8.0 \\
    \bottomrule
    \end{tabular}
    \end{minipage}
\end{table}

As summarized in Table~\ref{tab:duration_exposure}, \new{anomaly} events were typically short-lived: 71\% lasted less than 60 minutes with a median duration of 40 minutes. A small number, however, persisted for many hours, \new{potentially} indicating sustained or repeated interference. Per-vessel frequency was similarly skewed: 54\% of flagged ships experienced only one episode, whereas roughly 8\% exhibited twenty or more. These long-tail cases suggest recurrent exposure within stable interference zones or deliberate transponder manipulation.

\subsubsection{Maritime \new{Anomaly} Hotspots}
\label{sec:spoofing_hotspots}

\begin{figure}[]
  \centering
    \includegraphics[width=\linewidth]{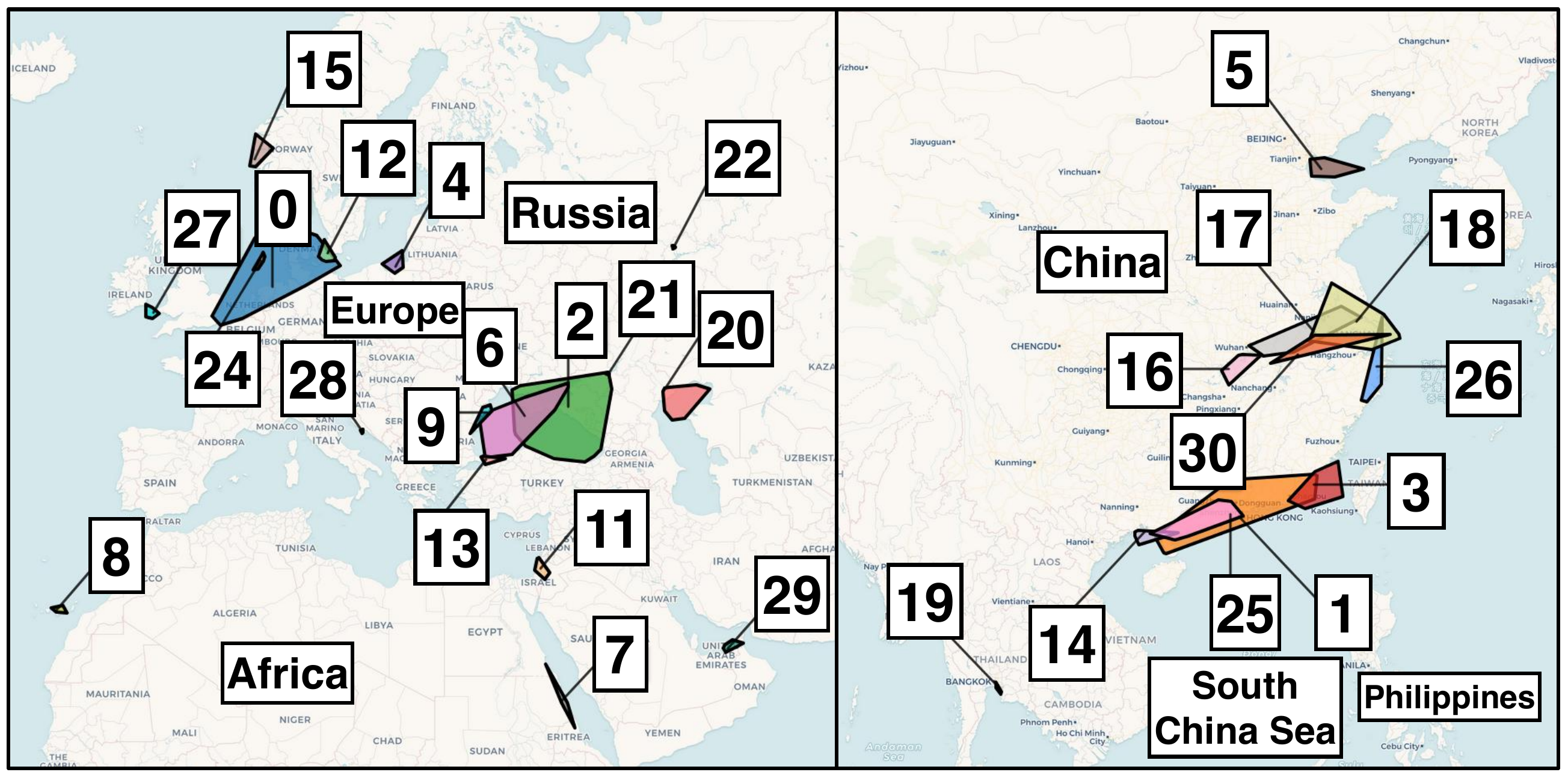}
    \caption{Global \new{anomaly} hotspots across Afro-Eurasia.}
  \label{fig:global_hotspots_afroeurasia}
\end{figure}
\begin{figure}[]
  \centering
    \includegraphics[width=\linewidth]{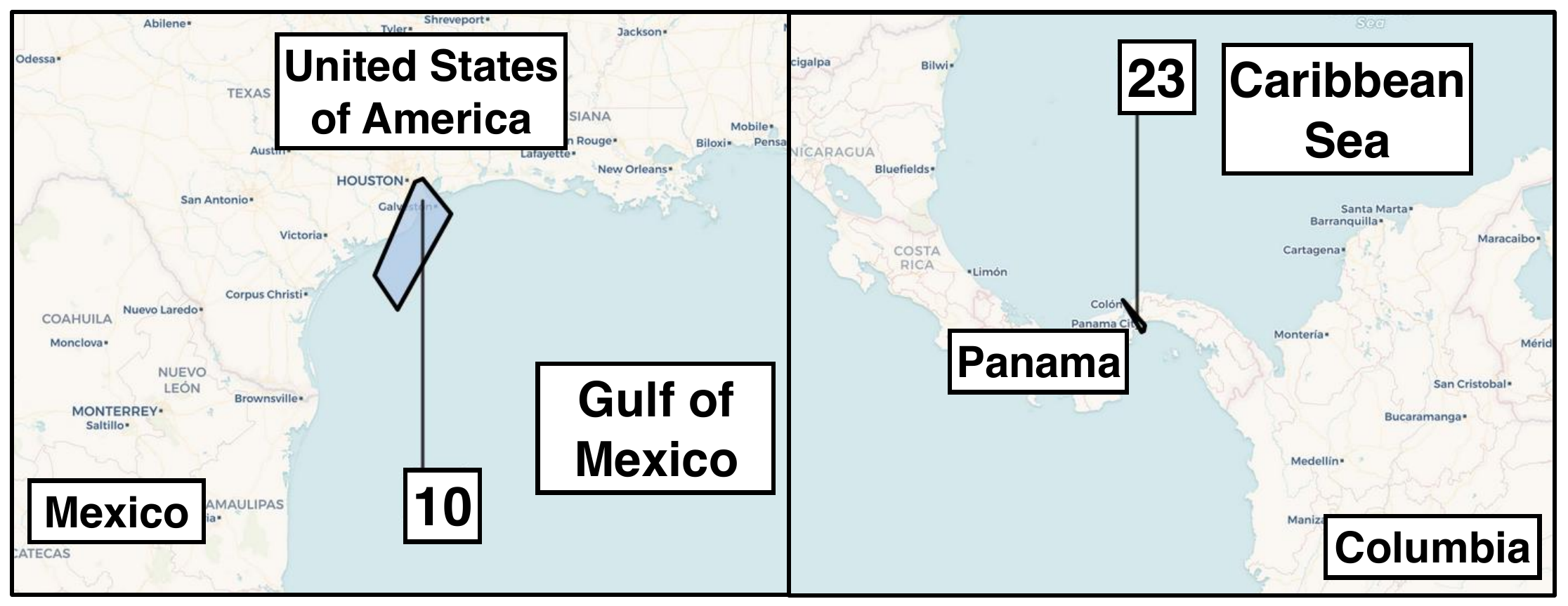}
    \caption{Global \new{anomaly} hotspots across the Americas.}
  \label{fig:global_hotspots_americas}
\end{figure}

\begin{table}[]
  \centering
  \caption{Regional summary of \new{anomaly} hotspots.}
  \label{tab:cluster_locations_short}
  \resizebox{0.90\columnwidth}{!}{%
  \small
  \setlength{\tabcolsep}{6pt}
  \renewcommand{\arraystretch}{1.2}
  \begin{tabularx}{\linewidth}{@{}p{0.56\linewidth} p{0.27\linewidth}@{}}
    \toprule
    \textbf{Region} & \textbf{Cluster IDs} \\
    \midrule
    Black Sea & 2, 6, 9, 13 \\
    Eastern Mediterranean \& Red Sea & 7, 11 \\
    North Sea \& Baltic & 0, 4, 12, 15, 24 \\
    East Asia Seas & 1, 3, 5, 14, 19, 25 \\
    China Inland Rivers & 16, 17, 18, 26, 30 \\
    Russia Inland Rivers & 21, 22 \\
    Caspian Sea & 20 \\
    Atlantic Corridors & 8, 27, 28 \\
    Gulf of Mexico \& Panama Corridor & 10, 23 \\
    Strait of Hormuz & 29 \\
    \bottomrule
  \end{tabularx}
  }
\end{table}

We next move from ships to space: aggregating \new{anomaly episodes} across vessels and time to locate global hotspots. Using the clustering framework described in Section~\ref{sec:part2}, we identify 31 persistent \new{anomaly} zones (Clusters 0-30) detected across more than two months of AIS traffic. Their regional distribution is summarized in Table~\ref{tab:cluster_locations_short}; the full ID-to-area map with coordinates is shown in Table~\ref{tab:cluster_location_map} in the appendix. Each cluster aggregates multiple independent vessels exhibiting spatially and temporally correlated spoofing-like motion.

To ensure robustness under varying \new{analysis thresholds}, we evaluate each cluster under two complementary thresholds: a lower bound, which captures high-confidence \new{anomaly} episodes, and an upper bound, which estimates the total potential exposure. The lower bound requires at least 30 minutes with $\ge$70\% anomalous points before confirming \new{an anomaly}. The upper bound includes any point violating the 60 knots speed, 5km kalman filter deviation, or on-land position constraints. Together, these bracket the plausible range of interference activity observed.

Figures~\ref{fig:global_hotspots_afroeurasia} and~\ref{fig:global_hotspots_americas} illustrate the global distribution of \new{anomaly} clusters visually, grouped into Afro-Eurasian and Pan-American regions. Most zones appear in geopolitically sensitive areas such as the Black Sea, Eastern Mediterranean, and South China Sea, aligning with regions previously linked to GPS interference and electronic warfare activity~\cite{c4ads_gps_spoofing_russia_syria,trevithick_gps_spoofing_china,harris_gps_mystery}. In contrast, several unexpected clusters emerged in civilian or low-risk regions, including the Canary Islands and Gulf of Mexico, where no prior public reporting of GPS spoofing exists. At their monthly peaks, the Canary Islands cluster affected 7 vessels under lower-bound criteria and 403 under the upper bound, while the Gulf of Mexico cluster affected 13 and 1,196 vessels, respectively. \new{Both this event and the Baltic cluster near Copenhagen (Cluster 12) were independently reproduced on public NOAA and Danish AIS data (Section~\ref{sec:dataset}), recovering the same patterns from open sources. These are the only clusters with sufficient coverage in the public feeds; coverage is too sparse for any other nearby clusters (e.g., Cluster 24), underscoring the need for Spire's global feed as our primary source.} Across all 31 clusters from Spire's feed, median \new{flagged-}vessel counts ranged from 22 to 1,852, spanning small transient events to large sustained zones.

\subsubsection{Recurring Geometric Signatures}
\label{sec:geometric}
Having mapped where \new{anomaly hotspots} occur globally, we next examine how they manifest spatially within each region. \new{The flagged} vessel trajectories reveal distinct geometric patterns that characterize different modes of interference. In both prior incident reports and our manual review of thousands of \new{flagged} trajectories, four recurring forms: circular, linear, convergent, and irregular, consistently appeared as the dominant modes of distortion~\cite{harris_gps_mystery,trevithick_gps_spoofing_china,gnss_msc_antonia,raza_msc_antonia,c4ads_gps_spoofing_russia_syria,arraf_israel_gps_spoofing}. These patterns capture recognizable ways in which spoofing alters reported motion, whether holding a vessel at a false point, shifting it along a line, or scattering it erratically, and can be systematically identified from AIS geometry. This list is not intended to be comprehensive; rather, it reflects the geometries documented across incident reports and observed in our dataset. Together, they reveal different operational behaviors of interference, from how easily activity can be detected in open-source data to whether ships were the intended targets at all. \new{Additionally, the feasibility of producing these patterns is established by documented real-world incidents: the same circular and linear geometries we observe have been independently confirmed as deliberate spoofing in the Black Sea, China, and the Strait of Hormuz~\cite{c4ads_gps_spoofing_russia_syria,trevithick_gps_spoofing_china,iran_spoofing}.}

\begin{figure}[]
  \centering
  \fbox{\includegraphics[width=0.80\linewidth]{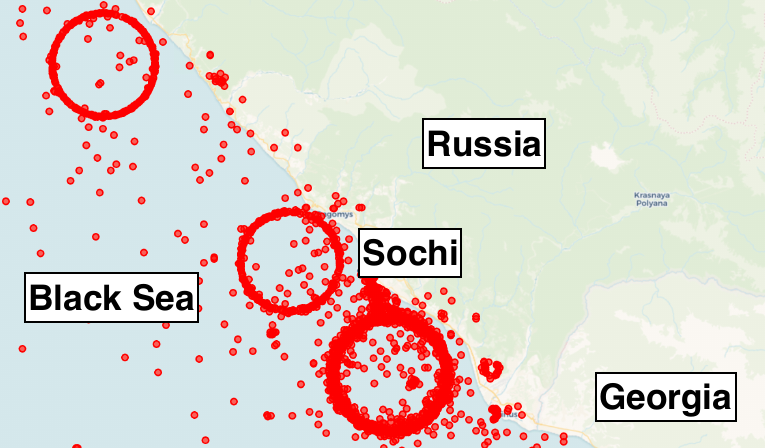}}
  \caption{Circular loops in the Black Sea region, showing multiple vessels exhibiting \new{spoofing-like} trajectories.}
  \label{fig:spoofing_circles}
\end{figure}

\begin{figure}[]
  \centering
  \fbox{\includegraphics[width=0.80\linewidth]{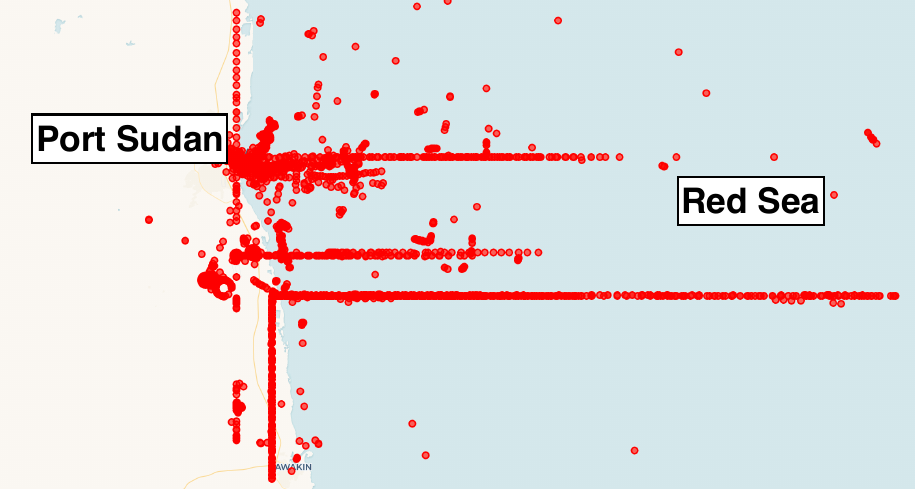}}
  \caption{Linear displacements in the Red Sea, showing the same lines that the \emph{MSC Antonia} Exhibited before grounding.}
  \label{fig:spoofing_lines}
\end{figure}

\noindent\textbf{1. Circular loops.}
Vessels appeared to travel in near-perfect circles, maintaining stable speed and course values while their reported positions rotated around a fixed point. We classified a cluster as circular when at least one vessel’s \new{anomalous} positions could be fitted to a circle with a radius between 0.2 km and 50 km, covering at least 180° of arc, and with $\ge$70\% of points lying close to that circle (mean radial deviation $\le$300 m or $\le$5\% of the radius). This behavior was most prominent in the Black Sea and the East China Sea/Chinese-coast regions, aligning with prior reports of ``crop-circle'' GPS spoofing in China and Russia~\cite{trevithick_gps_spoofing_china,c4ads_gps_spoofing_russia_syria}.

Circular trajectories are the most visually recognizable spoofing signature; they stand out clearly in vessel tracks and online mapping platforms, making detection easier for both analysts and casual observers~\cite{insidegnss_shanghai_spoofing}. The recurrence of these patterns in Russian and Chinese conflict-adjacent zones has given them a reputation as a hallmark of large-scale, state-linked interference~\cite{c4ads_gps_spoofing_russia_syria,harris_gps_mystery}. Their precision and frequency across many vessels suggest a persistent, coherent interference field rather than isolated noise~\cite{psiaki2016gnss}.

\noindent\textbf{2. Linear displacements.}
Vessels abruptly jumped from their true locations onto perfectly straight, uniform trajectories that extended for tens of kilometers before snapping back to their real positions. We classified these as linear when the \new{anomalous} positions aligned along a straight path $\ge$10 km in length with $\ge$75\% of points close to the line (orthogonal deviation $\le$250 m). The \emph{MSC Antonia} incident (Figure~\ref{fig:antonia_case}) exemplifies this pattern, where the vessel’s AIS track shifted onto a fixed-bearing line before grounding in the Red Sea. The defining feature of this pattern is its geometric regularity: straight, constant-bearing tracks that extend for tens of kilometers, a form of movement that is highly unnatural for real vessels and reflects deliberate falsification instead of noise~\cite{tippenhauer2011requirements,psiaki2016gnss}.

\noindent\textbf{3. Point convergence.}
In one region, multiple vessels simultaneously ``teleported'' to the same small inland area near Jordan’s Queen Alia Airport before returning to their true maritime positions. We classify a pattern as convergence when a compact area of radius $\le$300\,m contains at least 25\% of \new{anomalous} points from $\ge$5 vessels. Such collective displacement reflects broad-area interference affecting many receivers at once. Because the spoofed position is stationary and clearly impossible for ships, appearing on land at an airport, mariners would immediately recognize it as erroneous, indicating that vessels may be collateral recipients of interference directed at other systems. This behavior matches reports of regional anti-drone and air-defense GPS disruption~\cite{arraf_israel_gps_spoofing}, which pulls nearby receivers toward a single false fix.

\noindent\textbf{4. Irregular displacements.}
Some vessels exhibited scattered jumps or drifts within a confined area without forming stable geometric patterns. Clusters that failed the above criteria were labeled irregular, \new{consistent with} low-power or intermittent spoofing where false signals sporadically overpower authentic satellite reception~\cite{psiaki2016gnss}. Although less structured, these patterns are analytically significant: irregular trajectories are hardest to distinguish from benign anomalies such as AIS/GPS dropouts, MMSI reuse, or multipath reflections. Their presence underscores the need for cross-modal validation (e.g., speed consistency and cross-vessel timing) when inferring spoofing from open-source data.

\begin{table}[]
    \centering
    \caption{Clusters grouped by dominant geometric pattern, derived from vessel trajectories.}
    \label{tab:geometric_by_cluster}
    \begin{tabular}{l p{0.47\linewidth}}
    \toprule
    \textbf{Geometric Pattern} & \textbf{Clusters (IDs)} \\
    \midrule
    Circular loops & 0, 1, 2, 6, 7, 14, 16, 20, 21, 22, 25 \\
    Linear displacements & 3, 5, 8, 10, 12, 13, 17, 18, 19, 23, 24, 26, 29, 30\\
    Point convergence & 11 \\
    Irregular displacements & 4, 9, 15, 27, 28 \\
    \bottomrule
    \end{tabular}
    \end{table}

Figures~\ref{fig:spoofing_circles} and~\ref{fig:spoofing_lines} illustrate representative examples of the first two categories, (1) circular loops and (2) linear displacements, captured from the Black Sea and Red Sea, respectively. Together, these patterns demonstrate that \new{anomalies are} not uniform in manifestation: some zones maintain stable false coordinates for hours or even days, while others produce sporadic or partial positional distortions.

Table~\ref{tab:geometric_by_cluster} summarizes the dominant geometric pattern across all clusters. Linear displacements were the most prevalent, appearing in 14 of 31 clusters, while circular loops were observed in roughly one-third. This prevalence suggests that many interference sources generate simple, fixed-direction offsets that vessels interpret as straight-line motion, while a smaller subset produce stable circular trajectories that dominate receiver solutions for extended periods. Only one cluster (11, near Israel) exhibited a point-convergence pattern, which may be consistent with spillover from localized anti-drone or air-defense interference, and five clusters showed irregular trajectories indicative of weak or intermittent spoofing. Taken together, these four geometries capture the main operational ``shapes'' of \new{anomalous motion} we observe in AIS data.

\subsubsection{Temporal Persistence Patterns}
\label{sec:temporal}
Having characterized \new{anomalous} geometries, we next examine where and when these anomalies occur globally. Building on the 31 \new{anomaly} hotspots previously identified, we analyze how activity within these clusters evolves over time. Each zone represents a region where multiple vessels showed correlated spoofing-like motion over overlapping intervals.

\begin{figure}[]
  \centering
  \includegraphics[width=\linewidth]{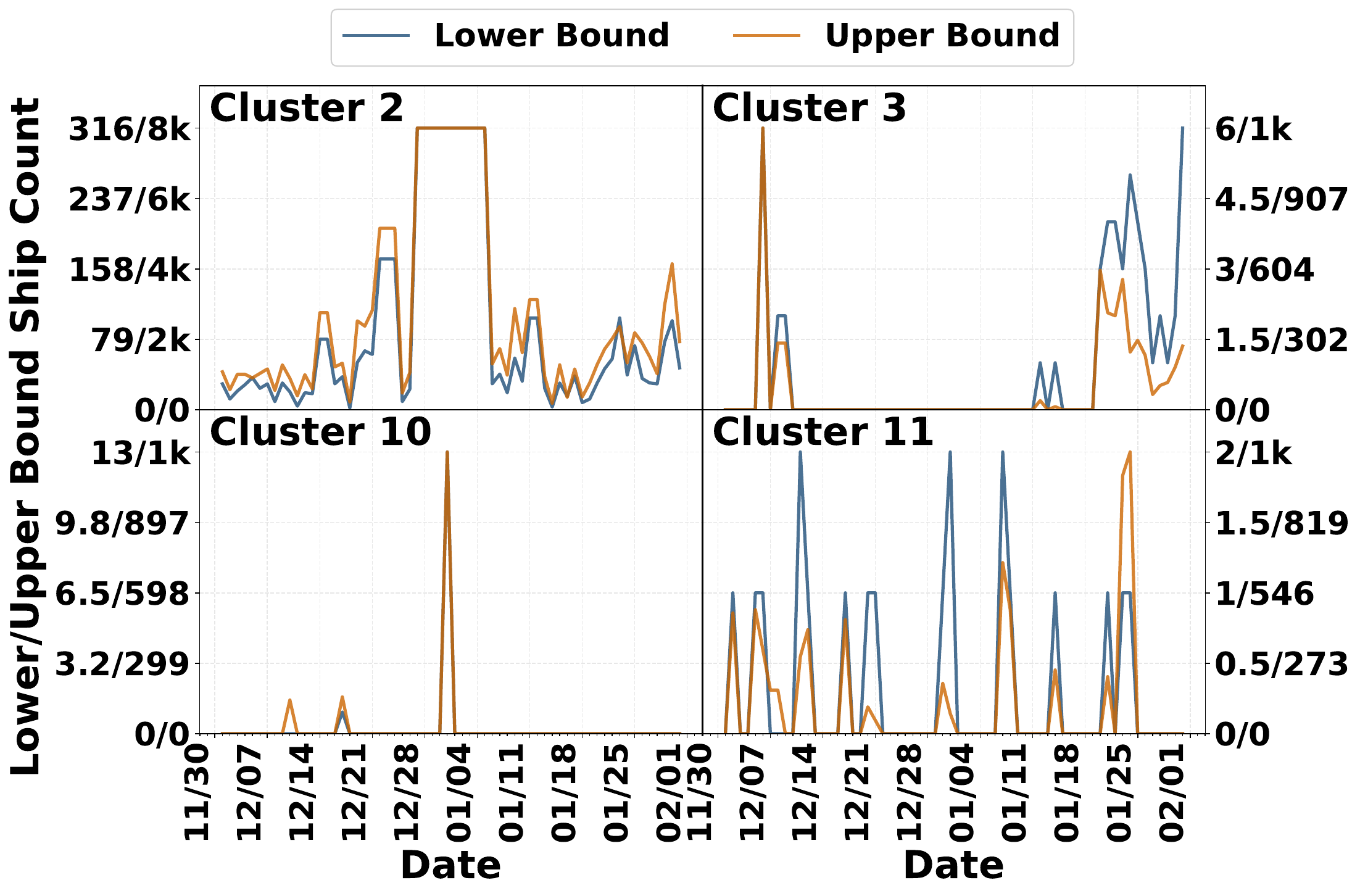}
  \caption{Representative temporal patterns:
(\textbf{top left}) sustained multi-day activity in the Black Sea (Cluster 2);
(\textbf{top right}) intermittent bursts in the Taiwan Strait (Cluster 3);
(\textbf{bottom left}) a single isolated event in the Gulf of Mexico (Cluster 10); and
(\textbf{bottom right}) recurrent \new{anomalies} off the Gaza coast (Cluster 11)}
  \label{fig:spoofing_timelines}
\end{figure}

\new{Anomaly} zones exhibit diverse temporal dynamics: some operate continuously for days, others reappear in bursts, and some vanish after brief episodes. For each cluster, we generated a one-minute-resolution time series representing the number of \new{flagged} vessels active at each point in time. This was computed by merging all overlapping \new{anomalous} intervals across ships within the same cluster, so that any period when multiple vessels \new{experienced} simultaneous \new{anomalies} appears as a single continuous interval.  Figure~\ref{fig:spoofing_timelines} illustrates the four temporal signatures observed across clusters: \\
\noindent\textbf{1. Sustained.}
Clusters were labeled as \emph{sustained} when vessel \new{anomalous} activity remained continuously elevated for at least seven consecutive days, with daily \new{flagged} vessel counts varying by no more than 10\%. This criterion captures long-duration interference zones characterized by stable intensity over time (Figure~\ref{fig:spoofing_timelines}, Cluster 2).

\noindent\textbf{2. Recurrent.}
Clusters were classified as \emph{recurrent} when the \new{flagged} vessel count exhibited at least three activity peaks separated by approximately regular intervals (coefficient of variation of inter-peak spacing $<$0.3). These clusters correspond to regions where \new{anomalies} reoccur on a periodic or scheduled basis (Figure~\ref{fig:spoofing_timelines}, Cluster 11).

\noindent\textbf{3. Intermittent.}
Clusters were labeled as \emph{intermittent} when they contained multiple irregular bursts of \new{anomalous} activity separated by quiet periods but did not meet the regularity or duration thresholds for sustained or recurrent behavior. This category captures sporadic or short-lived \new{anomaly} events (Figure~\ref{fig:spoofing_timelines}, Cluster 3).

\noindent\textbf{4. Isolated.}
Clusters were identified as \emph{isolated} when activity was dominated by a single short peak lasting fewer than three days, with all other peaks below 25\% of the maximum intensity. These represent one-off incidents or localized anomalies (Figure~\ref{fig:spoofing_timelines}, Cluster 10).

\begin{table}[]
    \centering
    \caption{Clusters grouped by temporal persistence patterns.}
    \label{tab:temporal_by_cluster}
    \begin{tabular}{{l p{0.49\linewidth}}}
    \toprule
    \textbf{Temporal Pattern} & \textbf{Clusters (IDs)} \\
    \midrule
    Sustained & 0, 2, 4, 5, 29\\
    Recurrent & 1, 11, 16, 22\\
    Intermittent & 3, 6, 7, 8, 12, 14, 17, 18, 20, 21, 25, 26, 30\\
    Isolated & 9, 10, 13, 15, 19, 23, 24, 27, 28\\
    \bottomrule
    \end{tabular}
    \end{table}

We classify all 31 \new{anomaly hotspots} identified in Figures~\ref{fig:global_hotspots_afroeurasia} and~\ref{fig:global_hotspots_americas} according to the empirical criteria described above. Table~\ref{tab:temporal_by_cluster} summarizes the dominant temporal persistence pattern for each cluster. Three broad trends emerge. First, sustained multi-day \new{anomalies} appear in the Black Sea, North Sea, and Strait of Hormuz, consistent with ongoing military spoofing activity~\cite{turgeon_russia_gps_spoofing}. \new{The Hormuz zone (Cluster 29) is notable for what came later: it appeared over a year before the 2026 Iran war drove mass GPS interference across the Persian Gulf~\cite{iran_spoofing}, underscoring the predictive value of persistent anomaly zones.} Second, recurrent and intermittent patterns dominate East Asian waterways and the Eastern Mediterranean, potentially reflecting localized and repeated interference events~\cite{poizner_gps_jamming}. Finally, shorter-lived and isolated incidents in the Gulf of Mexico and Panama Canal show that brief episodes occur even in regions without known conflict. These temporal signatures highlight both the persistence and diversity of global \new{anomaly} activity, showing that interference can be prolonged in conflict zones while remaining sporadic elsewhere.

\begin{figure}[]
  \centering
  \fbox{\includegraphics[width=\linewidth]{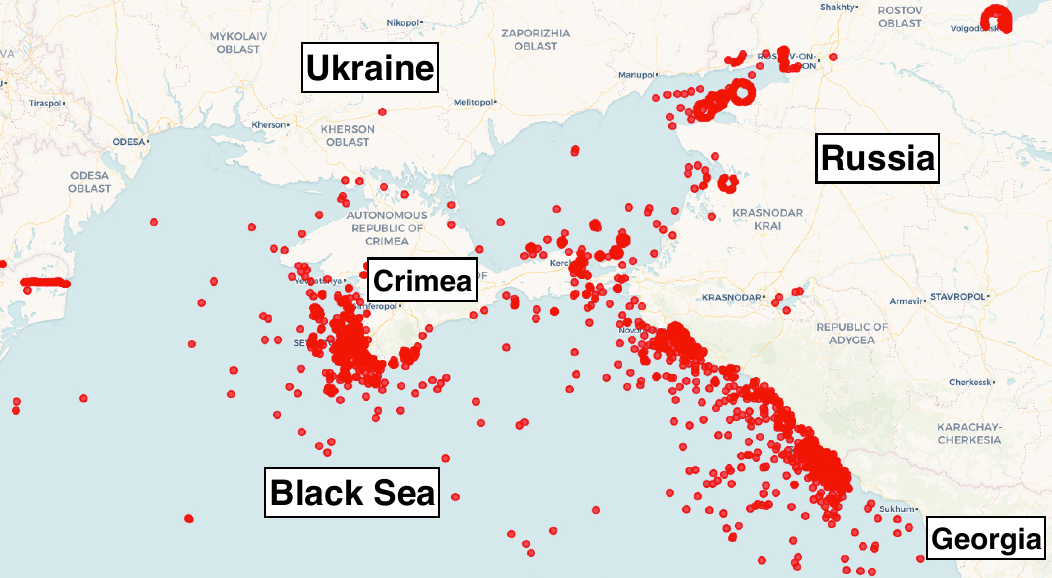}}
  \caption{\new{Anomalous} GPS points across the Black Sea. Dense clusters occur along the Russian and Crimean coasts.}
  \label{fig:blacksea_case}
\end{figure}

\begin{tcolorbox}[insightbox]
  \textbf{\new{Interference leaves fingerprints: anomaly zones recur with consistent geometries in the same places, on observable schedules. These zones can therefore be tracked over time, and in many regions, anticipated.}}
\end{tcolorbox}

\new{

\subsubsection{From Anomalies to Spoofing}
\label{sec:evidence_grading}
Now that we have identified 31 anomalous zones with characterized geometric and temporal patterns, we infer which zones are consistent with GPS spoofing. Recall that every zone has already passed the pipeline-level filtering of alternative explanations validated in Section~\ref{sec:validation} (Table~\ref{tab:alternatives-excluded}). What remains is to grade each zone by the strength of positive evidence supporting a spoofing interpretation.

\textbf{Grading positive evidence.} \new{We grade each remaining zone by the kind of evidence supporting a spoofing interpretation, defined by whether the spoofing is externally documented and how distinctive its signature is. Tier assignments follow the geometric and temporal criteria from Sections~\ref{sec:geometric} and~\ref{sec:temporal}, and were reviewed independently by multiple authors \cam{(not blinded)}, including a maritime-domain expert, who reached consensus on each label.}\\
\noindent\textbf{1. \new{Corroborated spoofing}}: \new{a distinctive spoofing signature together with external corroboration, through incident reports, fine-grained alignment with documented conflict events, or documented prior spoofing in the region (e.g., Black Sea crop-circles corroborated by multiple news reports). These are zones where independent evidence confirms spoofing is occurring.}\\
\noindent\textbf{2. \new{Newly identified spoofing}}: \new{a distinctive geometric or temporal signature without external corroboration (e.g., Gulf of Mexico synchronized linear jumps across roughly 200 vessels). Although no public report documents these zones, the signature itself is strong evidence: such geometries are implausible for genuine navigation and appear coherently across many independent vessels within a bounded region, the invariant pattern of regional external spoofing. These represent spoofing zones our measurement surfaces that have not, to our knowledge, been previously reported.}\\
\noindent\textbf{3. \new{Suggestive}:} \new{anomalies shared across vessels but with less distinctive signatures (e.g., inland river clusters and irregular displacement zones) and no external corroboration, consistent with spoofing but not individually conclusive.}

\begin{table}[]
\centering
\caption{Per-cluster evidence grading. Tiers reflect the strength of \textit{positive evidence} supporting GPS spoofing. \textit{Corroborated}: external corroboration. \textit{Newly Identified}: strong internal evidence without external corroboration. \textit{Suggestive}: weaker or less-structured internal evidence.}
\label{tab:evidence-grading}
\footnotesize
\setlength{\tabcolsep}{4pt}
\renewcommand{\arraystretch}{1.15}
\begin{tabular}{@{}p{2.2cm} p{1.8cm} p{3.6cm}@{}}
\toprule
\textbf{Region (IDs)} & \textbf{Tier} & \textbf{Positive Evidence} \\
\midrule
Black Sea \newline (2, 6, 9, 13) & Corroborated & Crop-circle patterns; C4ADS corroboration~\cite{c4ads_gps_spoofing_russia_syria}; conflict context \\
\addlinespace
Red Sea \newline (7) & Corroborated & Linear pattern; \textit{MSC Antonia} 5th-decimal match \\
\addlinespace
Gaza coast \newline (11) & Corroborated & Point convergence; daily alignment with conflict timeline \\
\addlinespace
Hormuz \newline (29) & Corroborated & Detected $>$1yr before Iran tensions; prior Iranian interference \\
\addlinespace
East Asia \newline (1, 3, 5, 14, 19, 25) & Corroborated & Crop-circle patterns; documented Shanghai/China spoofing \\
\midrule
Baltic \newline (0, 4, 12, 15, 24) & Newly identified & Sustained cross-vessel coherence; geometric regularity; high flag-state diversity \\
\addlinespace
Gulf of Mexico (10) & Newly identified & Cross-shaped jumps across $\sim$200 vessels; synchronized timing; high vessel diversity \\
\addlinespace
Panama Canal (23) & Newly identified & Coherent isolated-day geometry; cross-vessel consistency \\
\addlinespace
Canary Islands (8) & Newly identified & Recurrent pattern; cross-vessel coherence \\
\addlinespace
Caspian Sea (20) & Newly identified & Crop-circle pattern; consistent across vessels \\
\midrule
Inland rivers \newline (16, 17, 18, 21, 22, 26, 30) & Suggestive & Linear displacements; cross-vessel agreement; less flag diversity than other clusters \\
\addlinespace
Atlantic/Adriatic (27, 28) & Suggestive & Irregular patterns; cross-vessel correlation but less structured signatures \\
\bottomrule
\end{tabular}
\end{table}

\new{Of 31 zones, 13 are corroborated spoofing, 9 are newly identified spoofing supported by a strong internal signature, and 9 are suggestive. Detailed per-cluster grading appears in Table~\ref{tab:evidence-grading}. Under the most conservative interpretation, counting only corroborated zones, our results document spoofing across five geopolitical regions and more than 1,000 vessels; the newly identified tier extends this to previously unreported regions. The case studies that follow examine both corroborated zones and zones of distinct analytical interest, including unexpected civilian-zone anomalies and sanctions-related self-spoofing.}

\begin{tcolorbox}[insightbox]
  \textbf{\new{Across 31 anomaly zones, 22 show strong evidence of large-scale GPS spoofing. Of these, 13 align with documented incident reports, indicating our approach catches real, corroborated spoofing. The remaining 9 are \emph{previously undocumented}, showing our approach finds what existing sources miss.}}
\end{tcolorbox}

}

\subsection{Case Studies and Geopolitical Context}
\label{sec:case_studies}
Having established the global structure of spoofing activity, this section examines why these zones emerge and persist, integrating geopolitical, operational, and economic factors. We contextualize spoofing hotspots using conflict timelines, vessel attributes, and sanction status. Three representative examples capture both conflict-linked and civilian anomalies. A fourth case study explores how sanctioned vessels use self-spoofing to evade tracking (16-215 vessels under our bounds). These case studies corroborate that the detected clusters reflect real interference.

\noindent\textbf{Case Study 1: Black Sea and Russia.}
The Black Sea exhibited the most sustained and spatially widespread spoofing in our dataset, affecting between 902 and 2,297 vessels across both lower bound and upper bound thresholds. Interference concentrated around Sevastopol, Sochi, and Novorossiysk, extending eastward toward the Caucasus coast, while the northwestern shelf near Odesa and Constanța showed comparatively little activity. This east-west contrast suggests that spoofing is targeted within the region.

Circular ``crop-circle'' loops dominate nearly every active zone, sometimes persisting for several consecutive days.  
These loops are interspersed with abrupt, long-range jumps across the basin; linear displacements that project vessels onto false positions sometimes hundreds of kilometers away.  
Such geometry could indicate high-power, multi-receiver spoofing consistent with electronic-warfare~\cite{turgeon_russia_gps_spoofing,c4ads_gps_spoofing_russia_syria}.  

\begin{figure}[]
  \centering
  \fbox{\includegraphics[width=0.95\linewidth]{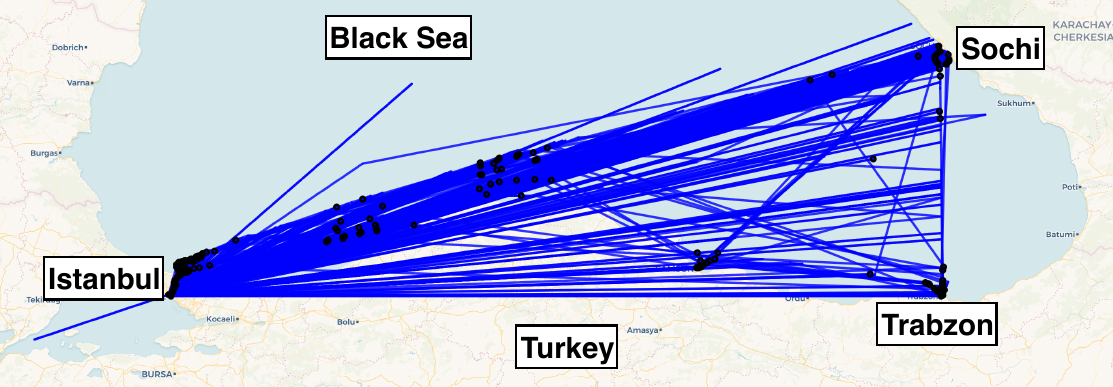}}
  \vspace{4pt}
  \fbox{\includegraphics[width=0.95\linewidth]{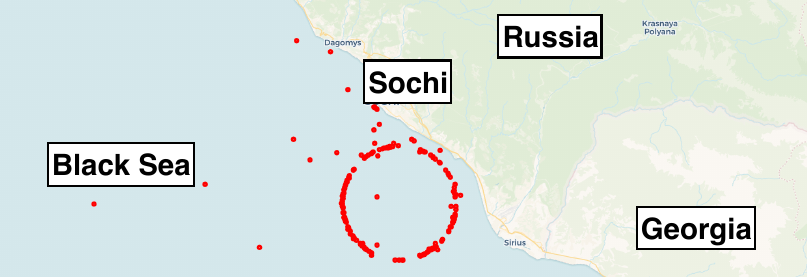}}
  \caption{Spoofing spillover on the civilian cruise ship \emph{Astoria Grande}.  
  (\textit{Top}) Distorted AIS track between Sochi, Istanbul, and Trabzon, showing unreadable routing from heavy interference.  
  (\textit{Bottom}) Circular spoofed points off Sochi, showing the ship was caught up in regional spoofing.}  
  \label{fig:astoria_case}
\end{figure}

Collateral impacts were also visible on civilian traffic.  
The passenger cruise ship \emph{Astoria Grande}, with a capacity of 1,699 passengers and crew~\cite{astoria_capacity}, was repeatedly spoofed into circular trajectories off Sochi, rendering its AIS record nearly unreadable (Figure~\ref{fig:astoria_case}). MarineTraffic data confirm its legitimate route between Sochi, Istanbul, and Trabzon~\cite{marinetraffic_astoria_grande,astoria_grande_route}, indicating that the interference was external to the vessel rather than self-induced. This spillover demonstrates how large-area spoofing operations can inadvertently endanger civilian vessels transiting conflict-adjacent waters.

\noindent\textbf{Case Study 2: Israel and the Red Sea.}
Spoofing activity \new{identified} across the Eastern Mediterranean and Red Sea exhibited both regional recurrence and distinct geometric signatures.  
Throughout the observation window, between 58 and 1,002 vessels (lower and upper bounds, respectively) were flagged for spoofing within this corridor, corresponding primarily to Clusters 7 and 11 from Figure~\ref{fig:global_hotspots_afroeurasia}.

\begin{figure}[]
  \centering
  \fbox{\includegraphics[width=0.80\linewidth]{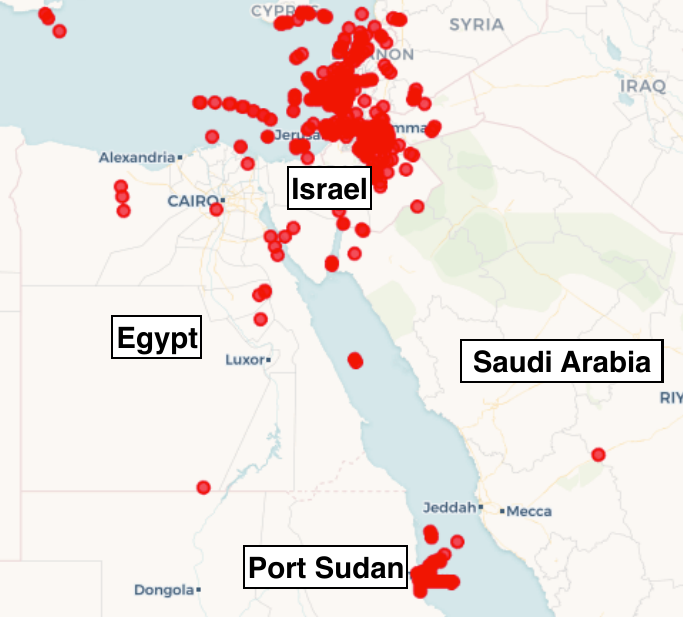}}
  \caption{Spoofed points across Israel and the Red Sea.  
  Linear displacements appear near Port Sudan, while dense convergence clusters occur inland near Israel and Jordan.}
  \label{fig:israel_red_sea}
\end{figure}

\begin{figure}[]
  \centering
  \vstretch{1}{\includegraphics[width=0.95\linewidth]
  {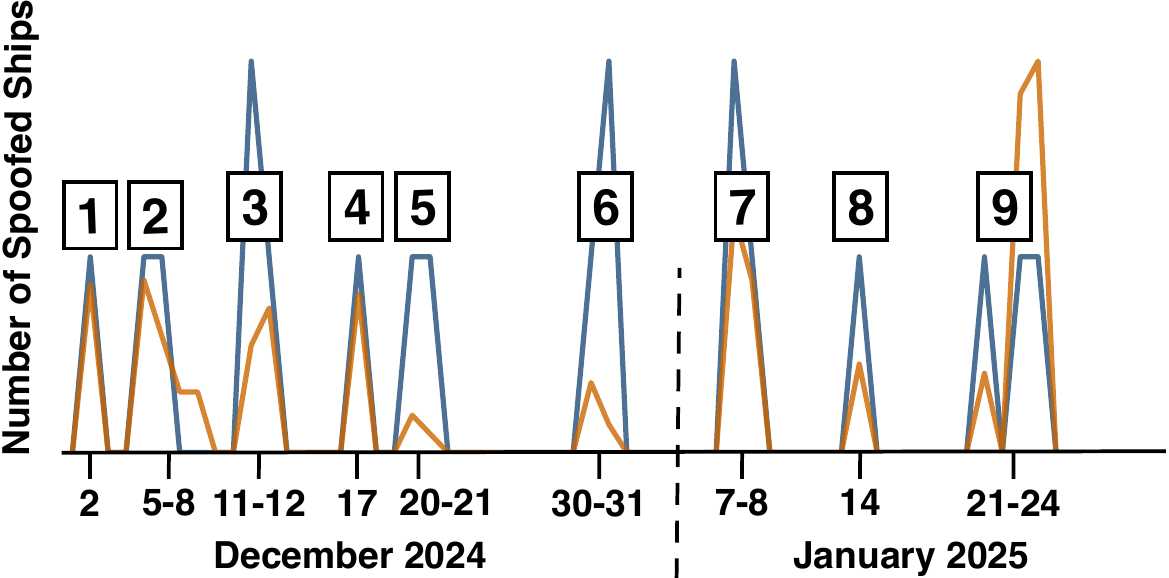}}
  \caption{Timeline of spoofing near Israel and Gaza, annotated with major conflict events (orange: upper bound; blue: lower bound); peaks align with reported escalations. 1: Ceasefire violated; 2: Gaza offensive; 3: Heavy strikes; 4: Ceasefire talks fail; 5: Houthi missile; 6: Year-end strikes; 7: Final Gaza strikes; 8: Day before ceasefire; 9: Post-truce raids}
  \label{fig:timeline_israel}
\end{figure}

Two dominant interference patterns emerge in the Israel-Red Sea corridor (Figure~\ref{fig:israel_red_sea}): point convergence and linear displacements. Near Gaza and the eastern Mediterranean coast, multiple vessels exhibited point convergence, where independent AIS tracks abruptly snapped to the same small inland coordinates at Queen Alia International Airport in Jordan. These convergence events were observed across dozens of vessels and occurred in repeated bursts that temporally aligned with major escalations in the Gaza conflict shown in Figure~\ref{fig:timeline_israel}.

These temporal overlaps are not isolated anecdotes. Independent timelines from Reuters, Al Jazeera, and Doctors Without Borders report similar peaks of hostilities, including heavy airstrikes (Dec~5–12), renewed raids (Dec~17–21), and major assaults in early January (Jan~7–8), which coincide with observed spoofing activity~\cite{aljazeera_timeline,reuters_timeline,doctors_timeline}. Together, these overlaps suggest that the spoofing was likely caused by air-defense activity, demonstrating how electronic-warfare during conflict can unintentionally disrupt civilian navigation and place vessels at risk.

Along the Red Sea corridor near Port Sudan, we observe recurrent linear displacements, parallel false tracks projecting tens of kilometers from the coast. These paths follow the exact same latitude, to the fifth decimal point, as those recorded from the \emph{MSC Antonia} prior to its grounding~(Figure~\ref{fig:antonia_case}), suggesting repeated or continued operation of the same interference source five months apart.

\begin{figure}[]
  \centering
  \fbox{\includegraphics[width=0.85\linewidth]{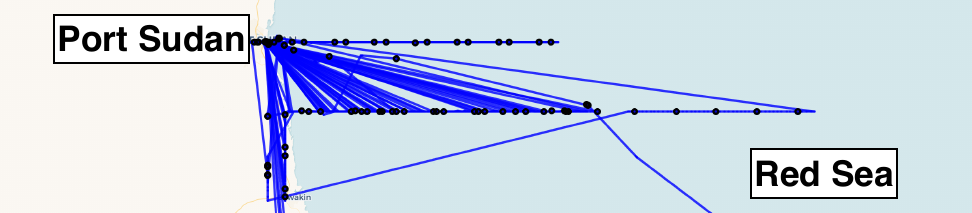}}
  \caption{Bulk carrier \emph{Charmous} with linear displacement in the Red Sea five months prior to \emph{MSC Antonia} grounding.}
  \label{fig:charmous_spoofing}
\end{figure}

The \emph{Charmous}, a bulk carrier flagged under Palau, provides a representative example.  
On January 21st 2025, its AIS track abruptly shifted into an artificial straight-line projection extending northeast from Port Sudan (Figure~\ref{fig:charmous_spoofing}).  
At least 11 other vessels (lower bound) recorded within 50 km of this event showed the same displacement at overlapping timestamps, suggesting an external interference source.

Collectively, these findings show that the Israel-Red Sea corridor features two recurring spoofing geometries: convergence near Israeli and Jordanian airspace, and linear displacements near Port Sudan. The Gaza-coast events consistently coincide with periods of active air-defense and drone operations, showing that spoofing traces may serve as useful signals for tracking regional military activity. In contrast, the repeated Red Sea displacements, first seen five months before the \emph{MSC Antonia} grounding (Figure~\ref{fig:antonia_case}), show that persistent spoofing zones can be monitored over time and may help warn mariners of navigation risks.

\begin{figure}[]
  \centering
  \fbox{\includegraphics[width=0.90\linewidth]{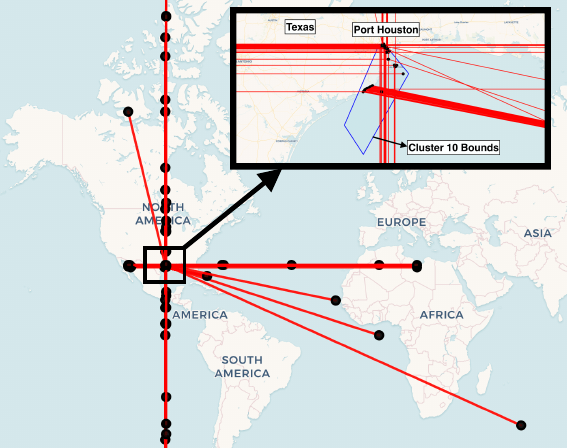}}
  \caption{Gulf of Mexico vessels (Cluster 10) show horizontal and vertical jumps, characteristic of GPS interference. \cam{Red lines indicate ship jump paths and black dots indicate the start and end locations of each jump.}}
  \label{fig:gulf_spoofing}
\end{figure}

\noindent\textbf{Case Study 3: Unexpected Spoofing Zones.}  
Beyond active conflict regions, we also identified spoofing in purely civilian maritime corridors, notably near the Canary Islands (Cluster 8), the Gulf of Mexico (Cluster 10), and the Panama Canal (Cluster 23). These short-lived, low-magnitude episodes lacked any connection to known conflicts or electronic-warfare. At their peaks, the Canary Islands cluster affected between 7 and 403 vessels, the Gulf of Mexico between 13 and 1,196, and the Panama Canal between 9 and 559 (lower-upper bounds). While the Gulf and Panama incidents appeared as isolated single-day events (December~31 and January~31, respectively), the Canary Islands exhibited recurrent spoofing throughout the observation period, including a sharp spike on December 31.

In the Gulf of Mexico, on December~31, almost 200 vessels exhibited synchronized positional jumps radiating north-south and east-west across thousands of kilometers (Figure \ref{fig:gulf_spoofing}). The uniform directions and simultaneous timing rule out multipath, random receiver error, \new{or the geomagnetic storm that struck later that day. \footnote{The storm and the jumps occurred at different times, and storm-induced errors would be far too small anyway~\cite{geomagnetic} (Section~\ref{sec:validation}).}} Instead, the observed geometry is more consistent with large-scale GPS spoofing, potentially originating from a coastal or airborne transmitter affecting multiple AIS receivers simultaneously. Given the region’s history of United States Navy and Coast Guard counter-narcotics and drone operations, this interference may reflect localized testing or operational GPS spoofing~\cite{insidegnss_texas_interference,hoagg_sirigu_dca_tcas}. \new{Furthermore, we reproduced the Gulf of Mexico event using public AIS data from NOAA~\cite{noaa_data}, recovering the same synchronized displacement pattern across 10 to 634 vessels (lower and upper bounds) on December 31. These track our Spire-based counts (13 and 1,196), with the lower NOAA figures reflecting its US-coastal terrestrial coverage versus Spire's global satellite feed. This independent result corroborates the finding and shows the pipeline generalizes across datasets.}

\noindent\textbf{Case Study 4: Sanctions and Self-Spoofing.} Cross-referencing spoofed vessel identifiers (MMSIs) with OFAC, EU, and UK sanctions lists revealed a small but significant subset of vessels, primarily oil tankers, linked to sanctioned entities~\cite{tankertrackers_blacklisted}. Several exhibited repeated spoofing or MMSI reuse along export corridors from Russian ports, showing they likely intentionally concealed their real locations.

\begin{figure}[]
  \centering
  \fbox{\includegraphics[width=0.80\linewidth]{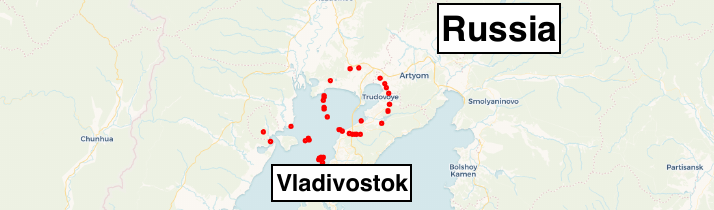}}
  \fbox{\includegraphics[width=0.80\linewidth]{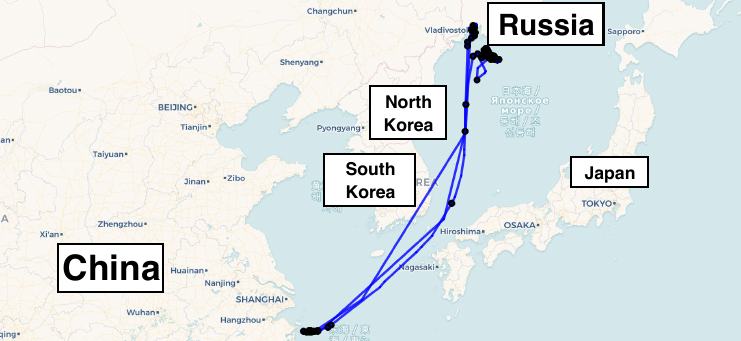}}
  \caption{Self-spoofing by the sanctioned tanker \emph{Hyperion}. (Top) Spoofed inland circular positions near Vladivostok. (Bottom) Route showing crude transport between Russian Pacific ports and China during the G7+ oil embargo period. }
  \label{fig:hyperion_selfspoof}
\end{figure}

The \emph{Hyperion}, was sanctioned by the U.S., EU, U.K., and others between January and July 2025 for transporting Russian oil using irregular, high-risk shipping practices. During our observation window, the vessel’s AIS data showed clusters of false inland positions near Vladivostok (Figure~\ref{fig:hyperion_selfspoof}), temporally aligned with its legitimate voyages to China. Other ships transiting the same corridor during the same period exhibited no comparable anomalies, indicating that the interference was localized to the vessel, consistent with self-spoofing.

A natural question is why a sanctioned vessel would falsify AIS positions rather than simply disable AIS. Prior work on sanctions-evasion notes that ``going dark'' is itself a major compliance red flag for insurers, ports, and classification societies, which routinely monitor AIS silence as a sign of illicit activity~\cite{bolton_sanctions}. Tankers engaged in sanctioned trade still require insurance and port access, making persistent outages more suspicious than misleading data~\cite{ihs_sanctions}. Manipulating coordinates, therefore, allows a vessel to maintain continuous transmission while obscuring its true movements.
Importantly, this self-spoofing is not meant to deceive nearby mariners or port authorities, who can physically observe the ship. The deception targets distant observers, satellite AIS receivers, commodity trackers, and regulators, whose global feeds can be misled during short falsified windows near loading or discharge sites~\cite{ihs_sanctions}. In \emph{Hyperion’s} case, spoofing was confined to port approaches: local actors would not be fooled, but remote monitors would record an altered movement history.

Beyond single-vessel manipulation, some sanctioned MMSIs were reused by other ships, and in other cases sanctioned vessels alternated between valid and falsified identifiers. This ambiguity, whether a sanctioned ship is hiding its location or being impersonated, complicates enforcement. Using our lower and upper bounds, we identify 16-215 sanctioned vessels engaged in likely self-spoofing~\cite{tankertrackers_blacklisted}, showing that GPS falsification has become a deliberate tool of sanctions evasion.

\begin{tcolorbox}[insightbox]
  \textbf{From conflict zones to commercial corridors, GPS spoofing has multiple motives: weapon, byproduct, and cover. In warzones it mirrors defense systems; at sea it entangles civilians; and along trade routes it conceals sanctioned movement. Together, these cases show how geopolitics shapes navigation integrity.}
\end{tcolorbox}

\section{Concluding Discussion}

Our analysis shows that GPS spoofing in maritime environments is neither rare nor random. Applying our two-stage framework to AIS data from late November 2024 to early February 2025, we identify 17,936 \new{anomaly} episodes across 2,663 vessels and 31 \new{anomaly} hotspots worldwide\new{, of which at least 22 show strong evidence of GPS spoofing (Sections~\ref{sec:spoofing_hotspots} and~\ref{sec:evidence_grading}). While many corroborate documented incidents, others are previously unreported spoofing regions that our measurement is the first to surface.} These zones cluster around conflict regions, major trade corridors, and sanctions-linked routes, and exhibit distinctive geometric (circular, linear, convergent, irregular) and temporal (sustained, recurrent, intermittent, isolated) signatures, as characterized in Sections~\ref{sec:geometric} and~\ref{sec:temporal}. In total, these findings demonstrate that maritime GPS interference is a structured, recurring feature of the global navigation environment, not merely isolated anomalies.\\
\noindent\textbf{Operational and Strategic Risks.}
Spoofing imposes operational hazards far beyond data integrity. Ships misled by falsified coordinates can drift into restricted waters, collide in dense corridors, or run aground, as exemplified by the \emph{MSC Antonia} incident, whose straight-line displacement matches a persistent Red Sea spoofing pattern (Figures~\ref{fig:spoofing_lines} and~\ref{fig:antonia_case}). Circular and linear geometries that appear as ``clean'' tracks over tens of kilometers are particularly dangerous: they are plausible enough to be followed, yet entirely divorced from the ship’s true position. In high-traffic regions such as the Eastern Mediterranean, Gulf of Mexico, and Panama Canal, even brief interference bursts can simultaneously affect hundreds of vessels (Figure~\ref{fig:gulf_spoofing}). At the strategic level, sustained and recurrent clusters in the Black Sea, Eastern Mediterranean, and East Asian waterways align with ongoing conflict and air-defense activity (Figures~\ref{fig:blacksea_case} and~\ref{fig:israel_red_sea}), while sanctions-linked tankers like the \emph{Hyperion} illustrate how self-spoofing serves regulatory evasion rather than purely military aims (Figure~\ref{fig:hyperion_selfspoof}).\\
\noindent\textbf{Human and Infrastructural Fragility.}
Modern ship navigation relies heavily on GPSes, leaving crews with few practical alternatives when spoofing occurs~\cite{noaa_paper_charts,ibaez2018TeachingCN}. Some spoofing geometries, such as vessels teleported onto an inland airport, are obvious to mariners; others, such as smooth straight-line drifts or irregular scatter, are much harder to detect in real time. Large commercial vessels require one to two nautical miles to stop or significantly alter course~\cite{wirz2012optimisation}, so even short delays in recognizing spoofing can translate into groundings or collisions, especially near shore where under-keel clearance is limited~\cite{navigator_ukc}. In practice, the skill and vigilance of mariners remain the last line of defense.\\
\noindent\textbf{From Measurement to Early Warning.}
A central insight from our study is that spoofing is \emph{measurable} and, in many regions, \emph{predictable}. Many hotspots exhibit stable or recurring temporal patterns, multi-day activity in the Black Sea and North Sea, recurrent bursts off Gaza, and intermittent episodes along Asian rivers and straits (Figure~\ref{fig:spoofing_timelines}, Table~\ref{tab:temporal_by_cluster}). \new{The Strait of Hormuz makes the predictive value concrete: our framework flagged sustained anomalies there in early 2025, over a year before the 2026 Iran war turned the same corridor into one of the most disrupted GPS environments in the world~\cite{iran_spoofing}. Had such monitoring been operational, the persistence of that zone would have been visible well in advance.} Because these interference zones reappear in the same locations with characteristic geometries, they can be monitored over time to produce risk maps akin to weather charts. Outputs from pipelines like ours (Section~\ref{sec:detection_and_characterization}) could be used for safety advisories or charting overlays, allowing ships to adjust routing and readiness before entering high-risk zones. For regulators, insurers, and coastal states, spoofing statistics provide intelligence on where navigation reliability is degraded and where enforcement or infrastructure hardening should be prioritized.\\
\noindent\textbf{Future Directions.}
Future work should extend this measurement over longer periods to capture seasonal and geopolitical variation, use additional sensors (e.g., radar, inertial logs) for stronger cross-validation, \new{and adapt the pipeline to run continuously as a near-live monitoring system.} Equally important is collaboration with maritime authorities and industry to translate measurement into operational tools such as alert services, dashboards, and risk indices usable by mariners with minimal additional training. Our results suggest that spoofing has evolved from isolated incidents into a persistent, organized feature of the maritime environment. Because many interference zones recur over time and exhibit distinctive geometries, continuous measurement enables early warning that could help prevent the next \emph{MSC Antonia}-like disaster.

\bibliographystyle{IEEEtran}
\bibliography{refs}

\begin{thebibliography}{100}
\providecommand{\url}[1]{#1}
\csname url@samestyle\endcsname
\providecommand{\newblock}{\relax}
\providecommand{\bibinfo}[2]{#2}
\providecommand{\BIBentrySTDinterwordspacing}{\spaceskip=0pt\relax}
\providecommand{\BIBentryALTinterwordstretchfactor}{4}
\providecommand{\BIBentryALTinterwordspacing}{\spaceskip=\fontdimen2\font plus
\BIBentryALTinterwordstretchfactor\fontdimen3\font minus
  \fontdimen4\font\relax}
\providecommand{\BIBforeignlanguage}[2]{{%
\expandafter\ifx\csname l@#1\endcsname\relax
\typeout{** WARNING: IEEEtran.bst: No hyphenation pattern has been}%
\typeout{** loaded for the language `#1'. Using the pattern for}%
\typeout{** the default language instead.}%
\else
\language=\csname l@#1\endcsname
\fi
#2}}
\providecommand{\BIBdecl}{\relax}
\BIBdecl

\bibitem{ics_shipping_data}
ICS, ``Shipping and world trade: World seaborne trade,'' 2025,
  \url{https://www.ics-shipping.org/shipping-fact/shipping-and-world-trade-largest-beneficial-ownership-countries/}.

\bibitem{macintyre_gps_spoofing_shipping}
I.~MacIntyre, ``The potentially catastrophic threat of gps spoofing in
  shipping,'' 2025,
  \url{https://www.imarest.org/resource/mp-the-potentially-catastrophic-threat-of-gps-spoofing-in-shipping.html}.

\bibitem{editorial_china_gps_spoofing}
{The Editorial Team}, ``Vessels navigating in china report gps spoofing
  incidents,'' 2020,
  \url{https://safety4sea.com/vessels-navigating-in-china-report-gps-spoofing-incidents/}.

\bibitem{emsa_2024}
EMSA, ``Annual overview of marine casualties and incidents 2024,'' 2024,
  \url{https://iims-media-library.s3.eu-west-2.amazonaws.com/wp-content/uploads/2024/12/17115213/Annual-Overview-of-Marine-Casualties-and-Incidents-2024.pdf#page=2.16}.

\bibitem{hancock_msc_antonia}
P.~Hancock, ``Msc antonia,'' 2025,
  \url{https://shipwrecklog.com/log/2025/05/msc-antonia/}.

\bibitem{lloyds_timeline}
R.~Willmington, ``Msc ship sails through bab el mandeb for first time since red
  sea exodus,'' 2025,
  \url{https://www.lloydslist.com/LL1154079/MSC-ship-sails-through-Bab-el-Mandeb-for-first-time-since-Red-Sea-exodus}.

\bibitem{wecox_antonia}
WEC, ``New casualty - msc antonia - grounding,'' 2025,
  \url{https://www.wecoxclaimsgroup.com/casualty-notices/new-casualty-msc-antonia-grounding/}.

\bibitem{lockton_antonia}
Lockton, ``Cyber-physical risk in the marine sector: a wake-up call from the
  msc antonia,'' 2025,
  \url{https://global.lockton.com/cca/en/news-insights/cyber-physical-risk-in-the-marine-sector-a-wake-up-call-from-the-msc-antonia}.

\bibitem{windward_front_eagle}
Windward, ``Gps jamming falsely placed vlcc front eagle in iran prior to
  collision,'' 2025,
  \url{https://windward.ai/blog/gps-jamming-falsely-placed-front-eagle-in-iran-prior-to-collision/}.

\bibitem{bockmann_uk_tanker}
M.~W. Bockmann, ``Seized uk tanker likely 'spoofed' by iran,'' 2019,
  \url{https://www.lloydslist.com/LL1128820/Seized-UK-tanker-likely-spoofed-by-Iran}.

\bibitem{chambers_msc_antonia}
S.~Chambers, ``Msc ship aground off jeddah, likely victim of gps spoofing,''
  2025,
  \url{https://splash247.com/msc-ship-aground-off-jeddah-likely-victim-of-gps-spoofing/}.

\bibitem{woody_navy_accidents}
C.~Woody, ``The navy's 4th accident this year is stirring concerns about
  hackers targeting us warships,'' 2017,
  \url{https://www.businessinsider.com/hacking-and-gps-spoofing-involved-in-navy-accidents-2017-8}.

\bibitem{raymaker2025sea}
A.~Raymaker, A.~Kumar, M.~Y. Wong, R.~Pickren, A.~Chhotaray, F.~Li, S.~Zonouz,
  and R.~Beyah, ``A sea of cyber threats: Maritime cybersecurity from the
  perspective of mariners,'' in \emph{ACM Conference on Computer and
  Communications Security (CCS)}, 2025.

\bibitem{tippenhauer2011requirements}
N.~O. Tippenhauer, C.~P{\"o}pper, K.~B. Rasmussen, and S.~Capkun, ``On the
  requirements for successful gps spoofing attacks,'' in \emph{ACM Conference
  on Computer and Communications Security (CCS)}, 2011.

\bibitem{zeng2018all}
K.~C. Zeng, S.~Liu, Y.~Shu, D.~Wang, H.~Li, Y.~Dou, G.~Wang, and Y.~Yang, ``All
  your gps are belong to us: Towards stealthy manipulation of road navigation
  systems,'' in \emph{USENIX Security}, 2018.

\bibitem{sathaye2022experimental}
H.~Sathaye, M.~Strohmeier, V.~Lenders, and A.~Ranganathan, ``An experimental
  study of gps spoofing and takeover attacks on uavs,'' in \emph{USENIX
  Security}, 2022.

\bibitem{balduzzi2014security}
M.~Balduzzi, A.~Pasta, and K.~Wilhoit, ``A security evaluation of ais automated
  identification system,'' in \emph{Proceedings of the 30th annual computer
  security applications conference}, 2014.

\bibitem{pavur2020tale}
J.~Pavur, D.~Moser, M.~Strohmeier, V.~Lenders, and I.~Martinovic, ``A tale of
  sea and sky on the security of maritime vsat communications,'' in \emph{IEEE
  Symposium on Security and Privacy (S\&P)}, 2020.

\bibitem{liu2021stars}
S.~Liu, X.~Cheng, H.~Yang, Y.~Shu, X.~Weng, P.~Guo, K.~C. Zeng, G.~Wang, and
  Y.~Yang, ``Stars can tell: a robust method to defend against gps spoofing
  attacks using off-the-shelf chipset,'' in \emph{USENIX Security}, 2021.

\bibitem{davidovich2022visas}
B.~Davidovich, B.~Nassi, and Y.~Elovici, ``Visas--detecting gps spoofing
  attacks against drones by analyzing camera's video stream,'' in \emph{Network
  and Distributed System Security Symposium (NDSS)}, 2022.

\bibitem{amro2022navigation}
A.~Amro, A.~Oruc, V.~Gkioulos, and S.~Katsikas, ``Navigation data anomaly
  analysis and detection,'' in \emph{Information}, 2022.

\bibitem{iran_spoofing}
D.~E. Béchard, ``Why ships in the strait of hormuz can’t trust their
  navigation screens,'' 2026,
  \url{https://www.scientificamerican.com/article/gps-spoofing-is-scrambling-ships-in-the-strait-of-hormuz/}.

\bibitem{imo_ais}
IMO, ``Ais transponders,''
  \url{https://www.imo.org/en/ourwork/safety/pages/ais.aspx}.

\bibitem{uscg_ais}
USCG, ``Automatic identification system (ais) overview,''
  \url{https://www.navcen.uscg.gov/automatic-identification-system-overview}.

\bibitem{lee2019maturity}
E.~Lee, A.~J. Mokashi, S.~Y. Moon, and G.~Kim, ``The maturity of automatic
  identification systems (ais) and its implications for innovation,''
  \emph{Journal of Marine Science and Engineering}, 2019.

\bibitem{tran2021marine}
K.~Tran, S.~Keene, E.~Fretheim, and M.~Tsikerdekis, ``Marine network protocols
  and security risks,'' in \emph{Journal of Cybersecurity and Privacy}, 2021.

\bibitem{progoulakis2021cyber}
I.~Progoulakis, P.~Rohmeyer, and N.~Nikitakos, ``Cyber physical systems
  security for maritime assets,'' in \emph{Journal of Marine Science and
  Engineering}, 2021.

\bibitem{direnzo2015little}
J.~DiRenzo, D.~A. Goward, and F.~S. Roberts, ``The little-known challenge of
  maritime cyber security,'' in \emph{International Conference on Information,
  Intelligence, Systems and Applications (IISA)}, 2015.

\bibitem{louart2023detection}
M.~Louart, J.-J. Szkolnik, A.-O. Boudraa, J.-C. Le~Lann, and F.~Le~Roy,
  ``Detection of ais messages falsifications and spoofing by checking messages
  compliance with tdma protocol,'' \emph{Digital Signal Processing}, 2023.

\bibitem{zheng2023identification}
H.~Zheng, Q.~Hu, C.~Yang, Q.~Mei, P.~Wang, and K.~Li, ``Identification of
  spoofing ships from automatic identification system data via trajectory
  segmentation and isolation forest,'' \emph{Journal of Marine Science and
  Engineering}, 2023.

\bibitem{androjna2021ais}
A.~Androjna, M.~Perkovi{\v{c}}, I.~Pavic, and J.~Mi{\v{s}}kovi{\'c}, ``Ais data
  vulnerability indicated by a spoofing case-study,'' \emph{Applied Sciences},
  2021.

\bibitem{androjna2023ais}
A.~Androjna, I.~Pavi{\'c}, L.~Gucma, P.~Vidmar, and M.~Perkovi{\v{c}}, ``Ais
  data manipulation in the illicit global oil trade,'' \emph{Journal of Marine
  Science and Engineering}, 2023.

\bibitem{bhatti2017hostile}
J.~Bhatti and T.~E. Humphreys, ``Hostile control of ships via false gps
  signals: Demonstration and detection,'' \emph{NAVIGATION: Journal of the
  Institute of Navigation}, 2017.

\bibitem{xu2023sok}
Y.~Xu, X.~Han, G.~Deng, J.~Li, Y.~Liu, and T.~Zhang, ``Sok: Rethinking sensor
  spoofing attacks against robotic vehicles from a systematic view,'' in
  \emph{IEEE European Symposium on Security and Privacy (EuroS\&P)}, 2023.

\bibitem{xu2024physcout}
Y.~Xu, G.~Deng, X.~Han, G.~Li, H.~Qiu, and T.~Zhang, ``Physcout: Detecting
  sensor spoofing attacks via spatio-temporal consistency,'' in \emph{ACM
  Conference on Computer and Communications Security (CCS)}, 2024.

\bibitem{clover_russia_gps_jamming}
C.~Clover and C.~Cook, ``How russia is jamming gps around europe,'' 2025,
  \url{https://www.ft.com/content/44cb37b6-2d82-402f-8c96-98b951f464af}.

\bibitem{gahnstrom_jamming_spoofing}
C.~J. Gahnström, ``Free article : Jamming and spoofing,'' 2024,
  \url{https://www.nautinst.org/resources-page/free-article-jamming-and-spoofing.html}.

\bibitem{spravil2023detecting}
J.~Spravil, C.~Hemminghaus, M.~von Rechenberg, E.~Padilla, and J.~Bauer,
  ``Detecting maritime gps spoofing attacks based on nmea sentence integrity
  monitoring,'' \emph{Journal of Marine Science and Engineering}, 2023.

\bibitem{jones_black_sea_spoofing}
M.~Jones, ``Spoofing in the black sea: What really happened?'' 2017,
  \url{https://www.gpsworld.com/spoofing-in-the-black-sea-what-really-happened/}.

\bibitem{zorri_pnt_weaponization}
D.~M. Zorri and G.~C. Kessler, ``Position, navigation, and timing weaponization
  in the maritime domain: Orientation in the era of great systems conflict,''
  2024,
  \url{https://ndupress.ndu.edu/Media/News/News-Article-View/Article/3678180/position-navigation-and-timing-weaponization-in-the-maritime-domain-orientation/}.

\bibitem{weill1997conquering}
L.~R. Weill, ``Conquering mutlipath: The gps accuracy battle,'' \emph{GPS
  world}, 1997.

\bibitem{esa_navipedia_multipath}
{European Space Agency}, ``Multipath,'' 2023,
  \url{https://gssc.esa.int/navipedia/index.php/Multipath}.

\bibitem{psiaki2016gnss}
M.~L. Psiaki and T.~E. Humphreys, ``Gnss spoofing and detection,''
  \emph{Proceedings of the IEEE}, 2016.

\bibitem{narain2019security}
S.~Narain, A.~Ranganathan, and G.~Noubir, ``Security of gps/ins based on-road
  location tracking systems,'' in \emph{IEEE Symposium on Security and Privacy
  (S\&P)}, 2019.

\bibitem{shen2020drift}
J.~Shen, J.~Y. Won, Z.~Chen, and Q.~A. Chen, ``Drift with devil: Security of
  multi-sensor fusion based localization in high-level autonomous driving under
  gps spoofing,'' in \emph{USENIX Security}, 2020.

\bibitem{yang2023location}
J.~Yang, A.~Estornell, and Y.~Vorobeychik, ``Location spoofing attacks on
  autonomous fleets,'' in \emph{Symposium on Vehicles Security and Privacy},
  2023.

\bibitem{tibaldo2025gnss}
C.~Tibaldo, H.~Sathaye, G.~Camurati, and S.~Capkun, ``Gnss-wasp: Gnss wide area
  spoofing,'' in \emph{USENIX Security}, 2025.

\bibitem{zhang2025ghost}
J.~Zhang, S.~Cheng, L.~Hu, J.~Zhang, C.~Shi, X.~Han, T.~Zhang, Y.~Cheng, and
  W.~Zhang, ``The ghost navigator: Revisiting the hidden vulnerability of
  localization in autonomous driving,'' in \emph{USENIX Security}, 2025.

\bibitem{sathaye2022semperfi}
H.~Sathaye, G.~LaMountain, P.~Closas, and A.~Ranganathan, ``Semperfi:
  Anti-spoofing gps receiver for uavs,'' in \emph{Network and Distributed
  System Security Symposium (NDSS)}, 2022.

\bibitem{jansen2018crowd}
K.~Jansen, M.~Sch{\"a}fer, D.~Moser, V.~Lenders, C.~P{\"o}pper, and J.~Schmitt,
  ``Crowd-gps-sec: Leveraging crowdsourcing to detect and localize gps spoofing
  attacks,'' in \emph{IEEE Symposium on Security and Privacy (S\&P)}, 2018.

\bibitem{coppola2025leo}
D.~Coppola, A.~Mumtaz, G.~Camurati, H.~Sathaye, M.~Singh, and S.~Capkun,
  ``Leo-range: Physical layer design for secure ranging with low earth orbiting
  satellites,'' in \emph{USENIX Security}, 2025.

\bibitem{cheng2025distributed}
X.~Cheng, H.~Yang, S.~Liu, and Y.~Yang, ``Distributed multi-antenna gps
  spoofing attack using off-the-shelf devices,'' in \emph{ACM Conference on
  Security and Privacy in Wireless and Mobile Networks}, 2025.

\bibitem{c4ads_gps_spoofing_russia_syria}
C4ADS, ``Above us only stars: Exposing gps spoofing in russia and syria,''
  2019,
  \url{https://c4ads.org/wp-content/uploads/2022/05/AboveUsOnlyStars-Report.pdf}.

\bibitem{harris_gps_mystery}
M.~Harris, ``Ghost ships, crop circles, and soft gold: A gps mystery in
  shanghai,'' 2019,
  \url{https://www.technologyreview.com/2019/11/15/131940/ghost-ships-crop-circles-and-soft-gold-a-gps-mystery-in-shanghai/}.

\bibitem{trevithick_gps_spoofing_china}
J.~Trevithick, ``New type of gps spoofing attack in china creates ``crop
  circles'' of false location data,'' 2019,
  \url{https://www.twz.com/31092/new-type-of-gps-spoofing-attack-in-china-creates-crop-circles-of-false-location-data}.

\bibitem{noaa_data}
{NOAA AIS Data},
  \url{https://coast.noaa.gov/htdata/CMSP/AISDataHandler/2024/index.html}.

\bibitem{dma_data}
{Danish Maritime Authority AIS Data}, \url{http://aisdata.ais.dk/}.

\bibitem{spire_global}
{Spire Global}, \url{https://spire.com/}.

\bibitem{spire_ca_contract_2024}
------, ``Spire global awarded ca \$1.41 million contract from government of
  canada for ship tracking data,'' 2024,
  \url{https://ir.spire.com/news-events/press-releases/detail/227/spire-global-awarded-ca-1-41-million-contract-from}.

\bibitem{wendel2001direct}
J.~Wendel, C.~Schlaile, and G.~F. Trommer, ``Direct kalman filtering of gps/ins
  for aerospace applications,'' in \emph{International Symposium on Kinematic
  Systems in Geodesy, Geomatics and Navigation (KIS2001)}, 2001.

\bibitem{alabsi2021tracking}
M.~Al-Absi, R.~Fu, K.~Kim, Y.-S. Lee, A.~Al-Absi, and S.~Lee, ``Tracking
  unmanned aerial vehicles based on the kalman filter considering uncertainty
  and error aware,'' \emph{Electronics}, 2021.

\bibitem{nishad2025advanced}
D.~K. Nishad, S.~Khalid, D.~Prakash, V.~K. Singh, and P.~Sahani, ``Advanced
  algorithms for uav tracking of targets exhibiting start-stop and irregular
  motion,'' \emph{Scientific Reports}, 2025.

\bibitem{levy1997kalman}
L.~J. Levy, ``The kalman filter: navigation's integration workhorse,''
  \emph{GPS World}, 1997.

\bibitem{itu_M.1371-5}
{International Telecommunication Union}, ``Recommendation itu-r m.1371-5,''
  2014,
  \url{https://www.itu.int/dms_pubrec/itu-r/rec/m/R-REC-M.1371-5-201402-S!!PDF-E.pdf}.

\bibitem{dbski_haversine}
{Kansas State University}, ``Distance between points on the earth’s
  surface,'' \url{https://www.math.ksu.edu/~dbski/writings/haversine.pdf}.

\bibitem{bansal2021deriving}
R.~Bansal, ``Deriving and testing the great circle theory,''
  \emph{International Journal of Statistics and Applied Mathematics}, 2021.

\bibitem{pekel2016high}
J.-F. Pekel, A.~Cottam, N.~Gorelick, and A.~S. Belward, ``High-resolution
  mapping of global surface water and its long-term changes,'' in
  \emph{Nature}, 2016.

\bibitem{jrc_dataset}
J.~R. Centre, ``Global surface water - data access,''
  \url{https://global-surface-water.appspot.com/download}.

\bibitem{cedelft_slow_steaming_2012}
J.~Faber, D.~Nelissen, G.~Hon, H.~Wang, and M.~Tsimplis, ``Regulated slow
  steaming in maritime transport,'' 2012,
  \url{https://theicct.org/wp-content/uploads/2021/06/CEDelft_slow_steaming_2012.pdf}.

\bibitem{uscg_rbm}
{United States Coast Guard}, ``Response boat–medium: Project profile,'' 2024,
  \url{https://www.dcms.uscg.mil/Our-Organization/Assistant-Commandant-for-Acquisitions-CG-9/Programs/Surface-Programs/Response-Boat-Medium/RBM-Profile-Copy/}.

\bibitem{usn_mkvi}
{U.S. Navy}, ``Mark vi patrol boat -- fact file,'' 2018,
  \url{https://www.navy.mil/Resources/Fact-Files/Display-FactFiles/Article/2173363/mark-vi-patrol-boat/}.

\bibitem{shaldag_mk2}
{Israel Shipyards}, ``Shaldag fast patrol craft (mk ii) - specifications,''
  2025, \url{https://www.israel-shipyards.com/naval-002.asp}.

\bibitem{noaa_weather}
NOAA, ``Space weather and gps systems,''
  \url{https://www.swpc.noaa.gov/impacts/space-weather-and-gps-systems}.

\bibitem{ester1996density}
M.~Ester, H.-P. Kriegel, J.~Sander, X.~Xu \emph{et~al.}, ``A density-based
  algorithm for discovering clusters in large spatial databases with noise,''
  in \emph{KDD}, 1996.

\bibitem{schubert2017dbscan}
E.~Schubert, J.~Sander, M.~Ester, H.~P. Kriegel, and X.~Xu, ``Dbscan revisited,
  revisited: why and how you should (still) use dbscan,'' \emph{ACM
  Transactions on Database Systems (TODS)}, 2017.

\bibitem{itu834}
\BIBentryALTinterwordspacing
{International Telecommunication Union}, ``Recommendation itu-r p.834-9:
  Effects of tropospheric refraction on radiowave propagation,'' 2017.
  [Online]. Available: \url{https://www.itu.int/rec/R-REC-P.834/en}
\BIBentrySTDinterwordspacing

\bibitem{wsc_top50}
{World Shipping Council}, ``The top 50 container ports,''
  \url{https://www.worldshipping.org/top-50-container-ports}.

\bibitem{lloyds_top100}
{Lloyd’s List}, ``One hundred container ports 2023,'' 2023,
  \url{https://www.lloydslist.com/one-hundred-container-ports-2023}.

\bibitem{gnss_msc_antonia}
{Inside GNSS}, ``Msc antonia grounding in the red sea attributed to suspected
  gnss spoofing,'' 2025,
  \url{https://insidegnss.com/msc-antonia-grounding-in-the-red-sea-attributed-to-suspected-gps-spoofing/}.

\bibitem{raza_msc_antonia}
R.~Raza, ``Container vessel msc antonia grounded in red sea; gps spoofing
  suspected,'' 2025,
  \url{https://www.marinetraffic.com/en/maritime-news/14/accidents/2025/12071/container-vessel-msc-antonia-grounded-in-red-sea-gps-spoofin}.

\bibitem{arraf_israel_gps_spoofing}
J.~Arraf, ``Israel fakes gps locations to deter attacks, but it also throws off
  planes and ships,'' 2024,
  \url{https://www.npr.org/2024/04/22/1245847903/israel-gps-spoofing}.

\bibitem{insidegnss_shanghai_spoofing}
{Inside GNSS}, ``Sinister spoofing in shanghai,'' 2019,
  \url{https://insidegnss.com/sinister-spoofing-in-shanghai/}.

\bibitem{turgeon_russia_gps_spoofing}
T.~Turgeon, ``Gps spoofing at russia’s borders: What to know,'' 2024,
  \url{https://www.gnssjamming.com/post/gps-spoofing-report-october-2024}.

\bibitem{poizner_gps_jamming}
S.~Poizner, ``From ukraine to taiwan, jamming of 50-year-old gps is a defense
  tech nightmare,'' 2024,
  \url{https://breakingdefense.com/2024/07/from-ukraine-to-taiwan-jamming-of-50-year-old-gps-is-a-defense-tech-nightmare/}.

\bibitem{astoria_capacity}
ShipCruises, ``Astoria grande,''
  \url{https://www.shipcruises.org/vessels/Astoria%20Grande/Ocean/10014}.

\bibitem{marinetraffic_astoria_grande}
{Marine Traffic}, ``Astoria grande,'' 2025,
  \url{https://www.marinetraffic.com/en/ais/details/ships/shipid:276298/mmsi:511100759/imo:9112789/vessel:ASTORIA_GRANDE}.

\bibitem{astoria_grande_route}
CruiseMapper, ``Astoria grande,'' 2025,
  \url{https://www.cruisemapper.com/ships/Astoria-Grande-657}.

\bibitem{aljazeera_timeline}
{Al Jazeera Staff}, ``Timeline: The path to the israel-hamas ceasefire deal in
  gaza,'' 2025,
  \url{https://www.aljazeera.com/features/2025/1/19/timeline-the-path-to-the-israel-hamas-ceasefire-deal-in-gaza}.

\bibitem{reuters_timeline}
Reuters, ``Israel-gaza war: A timeline of key events,'' 2025,
  \url{https://www.reuters.com/world/middle-east/major-moments-israel-gaza-war-2025-01-15/}.

\bibitem{doctors_timeline}
{Doctors without Borders}, ``Timeline: Bearing witness to genocide in gazan,''
  2024,
  \url{https://www.doctorswithoutborders.org/latest/timeline-bearing-witness-genocide-gaza}.

\bibitem{geomagnetic}
NOAA, ``G3 (strong) geomagnetic storm watch for 31 dec,'' 2024,
  \url{https://www.swpc.noaa.gov/news/g3-strong-geomagnetic-storm-watch-31-dec}.

\bibitem{insidegnss_texas_interference}
{Inside GNSS}, ``The unsolved mystery of the 2022 texas interference,'' 2023,
  \url{https://insidegnss.com/the-unsolved-mystery-of-the-2022-texas-interference/}.

\bibitem{hoagg_sirigu_dca_tcas}
A.~Hoag and G.~Sirigu, ``Dca tcas anomalies explained,'' 2025,
  \url{https://aireon.com/dca-tcas-anomalies-explained/}.

\bibitem{tankertrackers_blacklisted}
TankerTrackers, ``Officially blacklisted tankers,'' 2025,
  \url{https://tankertrackers.com/report/sanctioned/results}.

\bibitem{bolton_sanctions}
I.~Bolton, ``Sanctions at sea: Ais manipulation,'' 2025,
  \url{https://sanctionssos.com/myanmar-sanctions/f/sanctions-at-sea-ais-manipulation}.

\bibitem{ihs_sanctions}
I.~Markit, ``Sanctions advistories for the maritime industry,'' 2022,
  \url{https://library.iccwbo.org/content/tfb/pdf/AIS_Whitepaper_IHS_IIBLP_ACSS.pdf}.

\bibitem{noaa_paper_charts}
{NOAA}, ``Sunsetting traditional noaa paper charts,'' 2019,
  \url{https://nauticalcharts.noaa.gov/publications/docs/raster-sunset.pdf}.

\bibitem{ibaez2018TeachingCN}
I.~Iba{\~n}ez, ``Teaching celestial navigation in the age of gnss,''
  \emph{TransNav: International Journal on Marine Navigation and Safety of Sea
  Transportation}, 2018.

\bibitem{wirz2012optimisation}
D.-I.~F. Wirz, ``Optimisation of the crash-stop manoeuvre of vessels employing
  slow-speed two-stroke engines and fixed pitch propellers,'' \emph{Journal of
  Marine Engineering \& Technology}, 2012.

\bibitem{navigator_ukc}
{The Navigator}, ``All you ever wanted to know about under keel clearance…
  but were afraid to ask,'' 2021,
  \url{https://www.nautinst.org/resources-page/all-you-ever-wanted-to-know-about-under-keel-clearance-but-were-afraid-to-ask.html}.

\bibitem{anonymous_repo}
A.~Raymaker, ``Maritime spoofing,'' 2026,
  \url{https://github.com/annaraymaker/maritime_spoofing_measurement}.

\end{thebibliography}

\appendices
\section{Open Science}
In the interest of open science, we provide our \new{measurement} and clustering implementation in a repository, including documented thresholds, algorithm parameters, and code structure corresponding to the methodology described in Sections~\ref{sec:part1} and~\ref{sec:part2}~\cite{anonymous_repo}. \new{To support independent verification, our repository includes a fully reproducible pipeline on NOAA's public AIS data~\cite{noaa_data}, which recovers the Gulf of Mexico spoofing event, and the Danish Maritime Authority's feed~\cite{dma_data}, which reproduces the Baltic event from open data alone.}

\section{Additional Validation}
\label{sec:appendix_validation}
Complementing Section~\ref{sec:validation}, the only port with a slightly elevated flag rate, Hamburg (0.8\%), was attributable to MMSI reuse, where multiple ships broadcasting under one identifier produce spoofing-like jumps between distant locations. Manual investigation of representative false positives confirmed they stem from reused identifiers, cloned transmitters, or rare sensor-level GPS faults rather than genuine classification errors; such anomalies do not persist or recur across multiple vessels and are excluded during second-stage clustering. We classify these as ``faulty'' transmissions, distinct from spoofing.

\cam{\noindent\textbf{Parameter Sensitivity.} Table~\ref{tab:sensitivity} reports the sweep described in Section~\ref{sec:validation}. We set each range from outside our own data. The deviation threshold runs from 2\,km, just above the largest benign GPS error, to 10\,km, twice our value. MV runs from 50\% to 90\%. RW runs from 10 minutes, the shortest window that still holds several AIS reports, to 60 minutes. CGE runs from 30 to 360 minutes, well below and well above the 120-minute point where the observed gap distribution drops off sharply. The Kalman reset runs from 3 minutes, the longest reporting interval AIS mandates, to 12 minutes. We also recomputed spoofing zone recovery using 50\,km and 150\,km matching distances instead of 100\,km; the counts change by at most three zones, and usually not at all. The episode column shows that each parameter does take effect: changing the deviation threshold moves the episode count by a third, yet every zone and region is still recovered. Our published setting recovers all 31 zones by definition, since recovery is measured against it; what matters is how far each parameter can move before the recovered set changes. Overall, our findings are not sensitive to these choices: plausible values far from the ones we selected recover nearly the same zones and regions.}

\begin{table}[t]
\centering
\caption{\cam{Stage-1 parameter sensitivity. Each parameter is varied
independently with the others held at their published values (bold).
Episode counts show that each parameter takes effect; recovery is measured
geographically against the baseline run.}}
\label{tab:sensitivity}
\small
\setlength{\tabcolsep}{4pt}
\begin{tabular}{llrrrr}
\toprule
Parameter & Value & Episodes & Zones & Regions & Corrob. \\
 & & & /31 & /17 & /13 \\
\midrule
\multirow{5}{*}{Deviation (km)}
  & 2 & 19{,}401 & 31 & 17 & 13 \\
  & 3 & 19{,}240 & 31 & 17 & 13 \\
  & \textbf{5} & \textbf{17{,}936} & \textbf{31} & \textbf{17} & \textbf{13} \\
  & 7 & 16{,}399 & 31 & 17 & 13 \\
  & 10 & 14{,}571 & 31 & 17 & 13 \\
\midrule
\multirow{5}{*}{MV (\%)}
  & 50 & 26{,}427 & 29 & 17 & 12 \\
  & 60 & 21{,}827 & 29 & 17 & 12 \\
  & \textbf{70} & \textbf{17{,}936} & \textbf{31} & \textbf{17} & \textbf{13} \\
  & 80 & 13{,}772 & 27 & 15 & 12 \\
  & 90 & 7{,}680 & 17 & 10 & 7 \\
\midrule
\multirow{5}{*}{RW (min)}
  & 10 & 10{,}777 & 26 & 17 & 12 \\
  & 20 & 11{,}456 & 29 & 16 & 13 \\
  & \textbf{30} & \textbf{17{,}936} & \textbf{31} & \textbf{17} & \textbf{13} \\
  & 45 & 15{,}270 & 30 & 17 & 13 \\
  & 60 & 12{,}141 & 23 & 13 & 12 \\
\midrule
\multirow{5}{*}{CGE (min)}
  & 30 & 19{,}789 & 28 & 17 & 12 \\
  & 60 & 18{,}723 & 29 & 17 & 13 \\
  & \textbf{120} & \textbf{17{,}936} & \textbf{31} & \textbf{17} & \textbf{13} \\
  & 180 & 17{,}702 & 30 & 17 & 13 \\
  & 360 & 17{,}596 & 30 & 17 & 13 \\
\midrule
\multirow{4}{*}{Reset (min)}
  & 3 & 17{,}571 & 22 & 13 & 9 \\
  & 5 & 18{,}295 & 28 & 17 & 12 \\
  & \textbf{7} & \textbf{17{,}936} & \textbf{31} & \textbf{17} & \textbf{13} \\
  & 12 & 15{,}122 & 30 & 17 & 13 \\
\bottomrule
\end{tabular}
\end{table}

\begin{figure}[]
    \centering
    \includegraphics[width=0.47\textwidth]{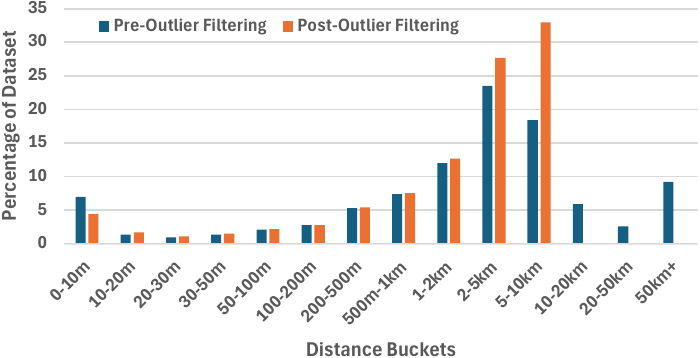}
    \caption{Error distribution across all AIS points before outlier filtering (left/blue) and after outlier filtering (right/orange).}
    \label{fig:error_data}
\end{figure}

\begin{figure}[]
    \centering
    \includegraphics[width=0.47\textwidth]{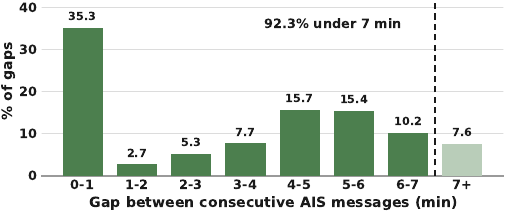}
    \caption{Distribution of time gaps between consecutive AIS points, motivating the 7-minute Kalman reset.}
    \label{fig:time_diff}
\end{figure}

\begin{figure}[]
  \centering
  \includegraphics[width=0.9\linewidth]{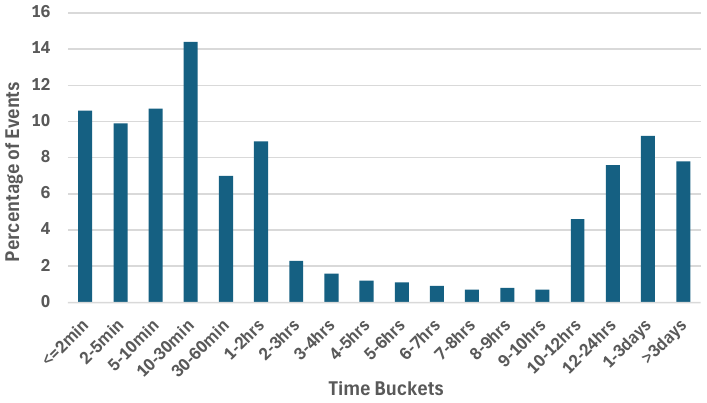}
  \caption{Distribution of clean-data gaps between events, motivating the 120-minute end threshold. }
  \label{fig:clean_gap}
\end{figure}

\begin{figure}[]
  \centering
  \includegraphics[width=\linewidth]{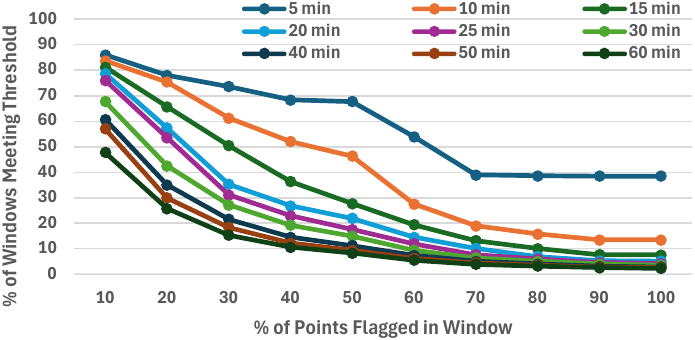}
  \caption{Empirical analysis of window size and anomaly ratio, justifying the 30min/70\% start threshold.}
  \label{fig:window_size}
\end{figure}

\begin{table}[]
\centering
\caption{Diversity and pattern consistency of detected spoofing clusters. High flag-state and vessel-type diversity, combined with shared spoofing geometries, makes coordinated self-spoofing unlikely.}
\label{tab:cluster_diversity}
\small
\begin{tabular}{lrrr}
\toprule
\textbf{Metric} & \textbf{Min} & \textbf{Median} & \textbf{Max} \\
\midrule
Unique flag states & 1 & 30 & 97 \\
Unique ship types & 2 & 5 & 10 \\
Dominant pattern share (\%) & 55 & 72 & 94 \\
\bottomrule
\end{tabular}
\end{table}

\begin{table*}[]
\caption{Validation of false-positives in high-traffic port regions; shows \new{pipeline} stability under dense GPS conditions.}
\label{tab:port_validation}
\resizebox{1.95\columnwidth}{!}{%
\centering
\begin{tabular}{|lccc | lccc|}
\hline
Port & \% Spoofed & Raw Spoofed & Total transited &
Port & \% Spoofed & Raw Spoofed & Total transited \\
\hline
Rotterdam       & 0.21\% & 28 & 13,073 & Antwerp         & 0.26\% & 12 & 4,656 \\
Singapore       & 0.01\% & 1  & 12,826 & Tianjin Xingang & 0.23\% & 6  & 2,577 \\
Guangzhou       & 0.05\% & 5  & 10,199 & Algeciras       & 0\%    & 0  & 2,243 \\
Shenzen         & 0.11\% & 9  & 7,906  & Piraeus Athens  & 0.05\% & 1  & 2,198 \\
Ningbo Zhoushan & 0.04\% & 3  & 6,814  & Kaohsiung       & 0.05\% & 1  & 2,077 \\
Busan           & 0.08\% & 5  & 6,265  & Qingdao         & 0\%    & 0  & 1,584 \\
Hong Kong       & 0.08\% & 5  & 5,900  & Hamburg         & 0.80\% & 11 & 1,373 \\
Tanger Med      & 0\%    & 0  & 4,910  & New York        & 0\%    & 0  & 1,292 \\
Los Angeles     & 0.09\% & 1  & 1,167  & Dubai           & 0\%    & 0  & 1,015 \\
Savannah        & 0.20\% & 1  & 493    & Klang           & 0\%    & 0  & 904   \\
Colombo         & 0\%    & 0  & 387    & Tanjung         & 0\%    & 0  & 729   \\
\hline
\end{tabular}
}
\end{table*}

\begin{table}[]
\caption{Cluster IDs mapped to geographic areas with approximate center coordinates.}
  \label{tab:cluster_location_map}
  \resizebox{0.95\columnwidth}{!}{%
  \centering
  \small
  \begin{tabularx}{\columnwidth}{@{}c p{0.55\linewidth} l@{}}
    \toprule
    \textbf{ID} & \textbf{Geographic area} & \textbf{Center (lat, lon)} \\
    \midrule
    0  & North Sea and Baltic corridor & (53.3934, 6.9707) \\
    1  & South China Sea near Hong Kong & (22.3627, 113.8641) \\
    2  & Eastern and northern Black Sea & (45.1837, 37.9976) \\
    3  & Taiwan Strait & (23.9420, 117.9925) \\
    4  & North Sea and Baltic corridor & (54.6332, 19.8848) \\
    5  & Yellow Sea & (38.6142, 118.3766) \\
    6  & Crimean sector of the Black Sea & (44.5062, 33.3763) \\
    7  & Red Sea near Port Sudan & (19.5410, 37.2871) \\
    8  & Canary Islands, off Western Sahara & (28.2174, -15.7286) \\
    9  & Western Black Sea & (44.7943, 29.4153) \\
    10 & Gulf of Mexico & (29.4044, -95.0533) \\
    11 & Eastern Med. off Gaza coast & (32.0352, 35.0947) \\
    12 & Baltic Sea near Copenhagen & (55.4358, 12.7245) \\
    13 & Southern Black Sea near Bosphorus & (41.1786, 29.3281) \\
    14 & South China Sea west of Hainan & (21.5156, 108.7343) \\
    15 & North Sea off Norway & (60.6318, 5.0696) \\
    16 & Yangtze River near Yueyang & (29.5911, 113.3469) \\
    17 & Yangtze River segment & (31.2860, 118.0208) \\
    18 & Yangtze River segment & (31.7975, 120.3062) \\
    19 & Gulf of Thailand & (13.5558, 100.6272) \\
    20 & Northern Caspian Sea & (46.3833, 48.0174) \\
    21 & Don River near Rostov Oblast & (47.5948, 42.1854) \\
    22 & Volga River near Kazan & (55.7906, 49.0417) \\
    23 & Panama Canal & (9.0218, -79.6625) \\
    24 & North Sea off Denmark & (54.9593, 5.7168) \\
    25 & South China Sea (open water) & (22.4166, 113.4405) \\
    26 & Yangtze River segment & (29.6902, 121.4221) \\
    27 & Celtic Sea & (51.7103, -5.5655) \\
    28 & Adriatic Sea & (43.3878, 16.3694) \\
    29 & Strait of Hormuz & (24.6216, 54.6840) \\
    30 & Yangtze River segment & (30.2764, 117.1088) \\
    \bottomrule
  \end{tabularx}
  }
\end{table}

\begin{table}[]
\centering
\caption{\cam{Stage-2 clustering parameter sensitivity. Each parameter is
varied with the other held at its published value (bold). Recovery is measured
geographically against the baseline run, as in Table~\ref{tab:sensitivity}.}}
\label{tab:dbscan-sensitivity}
\footnotesize
\setlength{\tabcolsep}{4pt}
\begin{tabular}{llrrr}
\toprule
Parameter & Value & Zones & Regions & Corrob. \\
 & & /31 & /17 & /13 \\
\midrule
\multirow{5}{*}{$\varepsilon$ (km)}
  & 25 & 22 & 15 & 9 \\
  & 40 & 27 & 17 & 12 \\
  & \textbf{50} & \textbf{31} & \textbf{17} & \textbf{13} \\
  & 75 & 26 & 17 & 10 \\
  & 100 & 19 & 16 & 7 \\
\midrule
\multirow{5}{*}{\textit{MinPts}}
  & 3 & 30 & 17 & 13 \\
  & 4 & 30 & 17 & 13 \\
  & \textbf{5} & \textbf{31} & \textbf{17} & \textbf{13} \\
  & 7 & 24 & 17 & 10 \\
  & 10 & 18 & 12 & 8 \\
\bottomrule
\end{tabular}
\end{table}

\section{Ethical Considerations}

\cam{\noindent\textbf{Stakeholders.}
Relevant stakeholders include mariners and passengers whose navigation may be affected by GPS interference; vessel owners and operators; the specific vessels discussed as case studies; ports, Vessel Traffic Services, coast guards, and navigation authorities; AIS data providers; governments and security organizations operating in regions where interference is observed; sanctions and maritime-enforcement authorities; researchers and maritime-security practitioners; and communities and industries that depend on safe commercial shipping. State or non-state actors conducting GPS interference are also stakeholders, because our analysis reveals that their activity is externally  observable.

\noindent\textbf{Respect for People.}
The study is passive and retrospective. We did not interact with vessels, crews, operators, navigation systems, or interference sources, and we did not transmit signals or alter maritime operations. AIS records identify vessels, but our analysis does not identify individual crew members, passengers, or other private persons. Results are primarily reported at the regional or aggregate level.

We use named-vessel examples when they are important to explain a finding or connect it to a documented event. We do not attribute responsibility to a vessel or crew when the evidence concerns external interference. Likewise, geographic or temporal association with an active conflict is treated as contextual evidence, not as attribution to any government, organization, or individual.

\noindent\textbf{Beneficence.}
The study is intended to improve understanding of a navigation-safety threat that already affects commercial shipping. Systematically identifying persistent interference can support mariner awareness, future warning systems, and research into more resilient navigation.

We also considered potential harms from reporting previously undocumented zones, including reputational harm, unsupported attribution, and revealing that an interference campaign is externally observable. We mitigate these risks by distinguishing Corroborated, Newly Identified, and Suggestive findings; avoiding attribution where the data cannot support it; focusing claims on observable interference patterns; and reporting findings primarily at the regional level. We do not release unnecessary per-vessel intermediate data for newly identified events.

\noindent\textbf{Justice.}
The risks and benefits of this work are not distributed uniformly. Mariners and commercial operators can be affected by regional GPS interference despite having no role in causing it, and vessels may appear in our dataset simply because they transit an affected region. We therefore treat affected vessels as potential victims or observers of interference, not as responsible parties, unless vessel-local manipulation is specifically supported by the evidence.

AIS coverage is also uneven. High-traffic regions and areas with better reception are more observable than remote maritime regions. We account for this as a measurement limitation and do not interpret an absence of detected activity as evidence that a region is free from interference.

\noindent\textbf{Respect for Law and Public Interest.}
The study analyzes passively collected AIS data obtained through a licensed
commercial provider and publicly available government sources. We did not bypass access controls, interfere with maritime systems, or perform active security testing.

The paper also distinguishes observation from attribution. Our data can support inference of regional GPS spoofing and characterize its observable effects, but generally cannot identify a transmitter or a responsible actor. This distinction is especially important for findings associated with geopolitical or military events. We believe publication serves the public interest by documenting a safety-relevant phenomenon from passive observations while avoiding unsupported attribution.

\noindent\textbf{Dual-Use Risk.}
We recognize that the pipeline has a dual-use dimension. An actor conducting GPS interference could use AIS data and our methodology to assess whether a campaign leaves an observable maritime footprint. However, the work does not provide instructions for generating counterfeit GPS signals, transmitting them, defeating receiver defenses, or making spoofing more effective. The dual-use information is primarily that sufficiently large regional interference can leave correlated signatures in vessel traffic.

We balance this risk against the safety value of allowing researchers and maritime operators to recognize persistent interference. Our reporting emphasizes regional patterns and does not provide unnecessary per-vessel intermediate outputs or operational detail beyond what is needed to support the scientific findings.

\noindent\textbf{Stakeholder Notification.}
We did not notify vessels, port authorities, the USCG, military organizations, or other operators. This study does not identify a specific exploitable vulnerability in a stakeholder's system, or an ongoing compromise that a particular organization can remediate. It retrospectively measures regional interference from observational data, and in many cases the data does not establish who generated the interference or which organization would be the appropriate disclosure recipient. Contacting an affected vessel could also incorrectly imply that the vessel or its operator was responsible for an externally induced anomaly, while contacting a suspected state or military actor would require an attribution that our evidence does not support. For these reasons, we treated this as a retrospective measurement study and did not conduct a vulnerability-disclosure process.}

\newpage %

\section{Meta-Review}

The following meta-review was prepared by the program committee for the 2027
IEEE Symposium on Security and Privacy (S\&P) as part of the review process as
detailed in the call for papers.

\subsection{Summary}
GPS spoofing has emerged as a common attack affecting location-dependent services. This study conducts a large-scale investigation of maritime GPS spoofing using global Automatic Identification System (AIS) data. The authors collected AIS data from Spire Global, NOAA, and the Danish Maritime Authority, covering more than 367,000 vessels. The authors develop a framework that allows for per-vessel anomaly detection based on unusual/infeasible location and movement data, and they apply anomaly-zone clustering to aggregate patterns and identify spoofed regions beyond individual vessel data.

\subsection{Scientific Contributions}
\begin{itemize}
\item Independent Confirmation of Important Results with Limited Prior Research
\item Creates a New Tool to Enable Future Science
\item Provides a Valuable Step Forward in an Established Field
\item Addresses a Long-Known Issue
\end{itemize}

\subsection{Reasons for Acceptance}
\begin{enumerate}
\item The study sheds new light on GPS spoofing in global AIS and vessel data.
\item The characterization of both spatial geometry and temporal persistence of candidate events is valuable.
\item The paper demonstrates a real-world threat based on real-world data with extensive experimentation
\end{enumerate}

\subsection{Noteworthy Concerns} %
\begin{enumerate} %
\item The reviewers found that the paper falls short of providing insights into maritime spoofing campaign operations and false-positive/missed-detection rates (due to missing ground-truth data).
\end{enumerate}

\section{Response to the Meta-Review}

We thank the reviewers for their positive assessment of our work. We would like to briefly address the two concerns raised in the meta-review.

\noindent\textbf{Spoofing campaign operations.} Our goal is to measure the prevalence, geographic distribution, and persistence of large-scale maritime GPS spoofing, rather than characterize the operational details of individual campaigns. Moreover, AIS data records the positions reported by vessels, not the underlying RF signals, making it impossible to determine the spoofing equipment, signal-generation techniques, or precise attacker objectives. We believe investigating these aspects is valuable future work, but it is outside the scope of this measurement study.

\noindent\textbf{Ground truth and detection accuracy.} To our knowledge, this is the first global-scale measurement of maritime GPS spoofing, and no corresponding ground-truth dataset exists. Consequently, we cannot directly measure false-positive or missed-detection rates. Instead, we validate our findings through independent AIS feeds, corroboration with documented incidents, checks against benign traffic, and extensive parameter sensitivity analyses. These experiments demonstrate that our main findings are robust to parameter choices, while we explicitly acknowledge the limitations of our detection approach.

We believe these limitations are inherent to the available measurement data. Our publicly released pipeline also enables future researchers to reproduce and extend our analysis.

\end{document}